\documentclass[12pt]{article}
\pdfoutput=1
\usepackage{amssymb}
\usepackage{amsmath}
\usepackage{graphicx}
\usepackage{hyperref}
\usepackage{bm}
\usepackage{color}

\ifx\affiliation\undefined 
\def\affiliation#1{\date{\normalsize #1\\ \today}}
\usepackage[margin=1in]{geometry}
\usepackage[sort&compress,numbers]{natbib}
\else 
\fi

\def\x{{\bm x}}

\def\hp{\hat\phi}
\def\tf{\tilde f}
\def\tF{\tilde F}

\def\sech{\,{\rm sech}}
\def\Real{{\rm Re}}\def\Imag{{\rm Im}}

\begin{document}
\title{Overstability of Plasma Slow Electron Holes}
\author{I H Hutchinson}
\affiliation{Plasma Science and Fusion Center, Massachusetts
  Institute of Technology, Cambridge, MA, USA} 

\ifx\altaffiliation\undefined\maketitle\fi 
\begin{abstract}
  Sufficient conditions are found on the ion velocity distribution
  $f_i$ and potential amplitude for stability of steady electron holes
  moving at slow speeds, coinciding with the bulk of
  $f_i$. Fully establishing stability requires calculation of the ion
  response to shift potential perturbations having an entire range of
  oscillatory frequencies, because under some conditions real frequencies
  intermediate between the ion and electron responses prove to be
  unstable even when the extremes are not. The mechanism of this
  overstability is explained and calculated in detail. Electron holes of
  peak potential $\psi$ less than approximately 0.01 times the background
  temperature ($\psi\lesssim 0.01T_0/e$) avoid the oscillatory
  instability entirely. For them, the \emph{necessary} condition that there
  be a local minimum in $f_i$ in which the hole resides is also
  \emph{sufficient}, unless the magnetic field $B$ is low enough to permit
  the transverse instability having finite wavenumber $k$
  perpendicular to $B$.
\end{abstract}
\ifx\altaffiliation\undefined\else\maketitle\fi  

\section{Introduction}

Plasma electron holes are solitary positive potential structures,
sustained by a deficit of collisionless electrons on trapped
phase-space orbits \citep{Hutchinson2017}. Most often, the velocity
$v_h$ of the frame in which the hole is steady moves much faster along
the magnetic field than typical ion velocities, with the result that
ion perturbation can be ignored. An extensive list of references to
satellite measurements of holes in various space regions is given in
the introduction of \cite{Lotekar2020}. However, \emph{slow electron
  holes}, defined as those for which there is a significant ion
population traveling at zero speed in the hole's rest frame, have also
recently been detected by satellite measurements in space plasmas
\citep{Graham2016,Steinvall2019,Lotekar2020,Kamaletdinov2021}. Past
simulations have almost always shown that slow holes do not
\emph{remain} slow, but are accelerated by repulsion from the
associated ion perturbation
\citep{Saeki1991,Muschietti1999,Eliasson2004,Eliasson2006,Zhou2016}
until they are fast, that is, their speed in the ion frame greatly
exceeds the ion velocity spread, and ion interaction has virtually
stopped. This poses a puzzle as to how slow electron holes can remain
slow, as they sometimes seem to do in nature.

That puzzle has recently been solved by showing that, although
single-humped ion velocity distributions always lead to the
self-acceleration instability, if a local minimum exists in the
distribution function $f_i(v)$, in which the hole velocity $v_h$ lies, then
the ion force on the hole reverses sign and becomes an effective
attraction \citep{Hutchinson2021c}. Moreover, in satellite measurements,
when slow holes are present, the background ion distributions are
observed to have just this feature: a local minimum at the hole
velocity \citep{Kamaletdinov2021}. Thus theory says that a necessary
condition for stability is that a local minimum in $f_i$ is present;
and in experiments this condition is satisfied.

What has not previously been shown theoretically, because the prior
analysis \citep{Hutchinson2021c} was essentially quasi-static, is
whether there are instabilities with non-zero real frequency (a
situation sometimes called overstability) remaining even if the
local minimum condition is satisfied. This is a significant concern,
because it is known that when an electron hole's velocity is within a
few ion sound speeds of even a narrow-velocity-spread ion distribution
whose phase-space density is negligible at the hole speed $v_h$, there
are oscillatory instabilities in hole
velocity \citep{Zhou2017,Zhou2018}. Moreover unpublished PIC simulations
(by the present author) of fully self-consistent slow electron holes
initialized in the presence of local $f_i$ minimum have shown slowly
growing velocity oscillations which eventually escape the attractive
force of the ion equilibrium charge, and become \emph{fast}.

The present theory therefore extends the previous analysis to a
full evaluation of the stability of shift perturbations of
arbitrary complex frequency. This requires one to use a fully
kinetic formulation of the responses of electrons \emph{and} ions,
evaluating the contribution of the entire velocity distributions,
abandoning any simplifying beam treatment of the ions. To my
knowledge such a full stability analysis has not previously been
accomplished. It is achieved by multidimensional numerical integration
of the perturbed force over the linearized perturbation solution of
the Vlasov equation in an electrostatic approximation.

In section \ref{sec2} the theoretical formulation and solution of the
linearized stability of slow electron holes is described. Section
\ref{sec3} gives one-dimensional stability results for equal ion and
electron temperatures, in which the mechanisms and regimes of
instability are identified. Section \ref{sec4} shows how the results
are affected by ion temperature, transverse wave-number, and magnetic
field strength.

\section{Formulation and Solution Method}\label{sec2}

Unless otherwise indicated, in this paper dimensionless units are used
with length normalized to Debye length
$\lambda_D=\sqrt{\epsilon_0T_0/n_0q_e^2}$, velocity to electron
thermal speeds $v_{te}=\sqrt{T_0/m_e}$, electric potential to thermal
energy $T_0/q_e$ and frequency to electron plasma frequency
$\omega_{pe}=v_{te}/\lambda_D$ (time normalized to
$\omega_{pe}^{-1}$).  This normalization reduces the one-dimensional
electron Vlasov equation to the form
\begin{equation}
  \label{eq:Vlasov}
  {\partial f\over \partial t}+v{\partial f\over \partial z}-
  q_e{\partial \phi\over\partial z}{\partial f\over \partial v} =0.
\end{equation}
The ion Vlasov equation is reduced to exactly the same form by using
units in which time is measured instead in terms of
$\omega_{pi}=\sqrt{m_e/m_i}\;\omega_{pe}$, except that instead of
$q_e=-1$ the particle charge for ions is simply reversed $q_i=1$ (so
$q\phi$ is the potential \emph{energy}). Ions in this paper all have
mass $m_i=1836m_e$; they are protons.

The formulation and analysis of the electron response follows closely
the treatment of previous papers
\citep{Hutchinson2018a,Hutchinson2019,Hutchinson2019a}, which should
be consulted for the mathematical derivation. Only an outline
description of the method is given here (thus glossing over a large
amount of algebraic and numerical work). Vlasov's equation can be
integrated along the equilibrium (zeroth order) orbits
to obtain the first-order perturbation to the distribution function
$f_1$ caused by a first-order electric potential perturbation
(relative to the non-uniform hole equilibrium)
\begin{equation}
  \label{eq:eigenmode}
  \phi_1(\x,t)=\hp(z)\exp i(ky-\omega t),  
\end{equation}
(including a perturbation transverse wave vector in the $y$-direction,
which for the first three sections of the paper will simply be $k=0$, i.e.\
one-dimensional variation along the applied uniform magnetic field
direction $z$).  The integration along unperturbed helical orbits
leads to an expansion in (integer $m$) harmonics
$\omega_m=m\Omega+\omega$ of the cyclotron frequency ($\Omega=eB/m_e$,
which represents the magnetic field strength), in which the
potential orbit-integrated over earlier time is
\begin{equation}
  \label{eq:phim}
  \Phi_m(z,t)\equiv 
  \int_{-\infty}^t \hp(z(\tau)){\rm e}^{-i\omega_m(\tau-t)}d\tau,
\end{equation}
where $z(\tau)=z(t)+\int_t^\tau v_z(t')dt'$ is the orbit position at earlier
time $\tau$. For positive imaginary part of $\omega$
the integrals converge and the perturbation in the parallel
distribution function (integrated over Maxwellian transverse
velocities) then can be found as \citep[see][eq:5.9]{Hutchinson2018a}
\begin{eqnarray}\label{eq:f1magnetic}
  f_{e\parallel 1}(y,t) =  
  q_e\phi_1(t)\left.{\partial f_{e\parallel0}\over\partial W_{e\parallel}}\right|_t
  + \sum_{m=-\infty}^\infty i\left[\omega_m
  {\partial f_{e\parallel0}\over \partial W_{e\parallel}}
  +(\omega_m-\omega) {f_{e\parallel0}\over T_\perp}\right]
  q_e\Phi_m {\rm e}^{-\zeta_t^2}I_m(\zeta_t^2)
  {\rm e}^{i(ky-\omega t)},
\end{eqnarray}
where $q_e$ is electron charge ($=-1$ in non-dimensionalized units),
$W_{e\parallel}$ is the parallel energy ${1\over 2}v_z^2+q_e\phi$
($m_e$ being unity), $f_{e\parallel0}$ is the unperturbed parallel
distribution function, a function only of energy, and $T_\perp$ is the
perpendicular temperature. The parameter $\zeta_t$ is the transverse
wavenumber $k$ times the thermal Larmor radius, so that
$\zeta_t^2=k^2T_\perp/\Omega^2$, and $I_m$ is the modified Bessel
function. When $\zeta_t$ is small, e.g.\ because $k$ is small (zero
even) or $\Omega$ is large, then the small-argument Bessel function
behavior ($I_m(\xi)\to (\xi/2)^m/\Gamma(m+1)$) implies that only the
$m=0$ cyclotron harmonic is non-negligible. It corresponds to a
one-dimensional motion treatment.

The first term of eq.\ (\ref{eq:f1magnetic}) is the ``adiabatic''
perturbation that would arise from a steady potential difference. It
does not contribute to the needed response. The remaining harmonic sum
is the non-adiabatic contribution denoted $\tf_{e\parallel}$.

A key formal difficulty is to find the shape of the linearized
eigenfunction $\hp(z)$ which self-consistently satisfies the Vlasov
and Poisson equations; this is an integro-differential
eigenproblem. For slow time dependence relative to particle transit
time, it can be argued on general grounds that the eigenmode consists
of a spatial shift (by small distance $\xi$ independent of
position) of the equilibrium potential profile ($\phi_0$)
 \citep[see][section 3.1]{Hutchinson2018a} giving:
\begin{equation}
  \label{eq:shiftmode}
  \hp = - \xi {\partial \phi_0\over \partial z},
\end{equation}
which is the perturbation we analyze.  Although this is only an
approximation to the exact eigenmode, the corresponding eigenvalue of
our system can be found to next order accuracy by using a ``Rayleigh
Quotient'' variational approximation \citep[see e.g.][]{Parlett1974}.
This mathematical procedure is equivalent to requiring the total force
exerted on the particles by the electrostatic field under the
influence of the assumed shift eigenmode to equal zero (or to balance
the transverse momentum transfer by Maxwell stress if $k\not=0$,
\citep[see][ section 3]{Hutchinson2018a}). It is the kinematic
momentum balance of the electron hole regarded as a rigid composite
object.  Suppressing the linear $\xi$ factor, the force on the
electrons is equal to the force of $\phi_0$ on the non-adiabatic
perturbed electron density
\begin{equation}
  \label{electronforce}
  \tF_e=  -\int{d\phi_0\over dz}q_e\left( \int \tf_{e\parallel} dv_z\right)   dz.
\end{equation}
The transverse Maxwell stress force when $k\not=0$ is 
\begin{equation}
  \label{eq:forcebalance}
  F_E \equiv -k^2 \int {d\phi_0\over dz}\hat\phi\, dz ,
\end{equation}
into which we substitute the shift-mode, eqs.\ (\ref{eq:shiftmode} and
\ref{eq:eigenmode}). 

The essential departure from the previous analysis, required here
for \emph{slow} electron holes, is that the force on the ions cannot be
ignored. By the same analysis as the electrons, but substituting ion
parameters, that force is 
\begin{equation}
  \label{ionforce}
  \tF_i=  -\int{d\phi_0\over dz}q_i\left( \int \tf_{i\parallel} dv_z\right) dz,
\end{equation}
where
\begin{eqnarray}\label{eq:fimagnetic}
  \tf_{i\parallel} = \sum_{m=-\infty}^\infty i\left[\omega_{im}
  {\partial f_{i\parallel0}\over \partial W_{i\parallel}}
  +(\omega_{im}-\omega) {f_{i\parallel0}\over T_{i\perp}}\right]
  q_i\Phi_m {\rm e}^{-\zeta_{it}^2}I_m(\zeta_{it}^2),
\end{eqnarray}
with $\Omega_i=(m_e/m_i)\Omega$, $\omega_{im}=\omega+m\Omega_i$, and
$\zeta_{it}^2=k^2T_{i\perp}/\Omega_i^2$. 
In practice, it is most convenient to calculate the ion force by the
same code as the electron force, except reversing the potential sign
(as proxy for charge sign), and using a frequency ($\omega$) scaled to
be larger by $\sqrt{m_i/m_e}$. Effectively that amounts to solving the
ion perturbation in \emph{ion} scaled units, in which the length scale is the
same (using the same reference temperature) and so for the same
density we get the same scaled force units.

The specific hole equilibrium potential shape analyzed is chosen to be
\begin{equation}
\label{holeequil}
  \phi_0 = \psi \sech^4(z/4).
\end{equation}
The background (untrapped) electron parallel velocity distribution is
taken as an unshifted Maxwellian of temperature $T_0$, and
$T_\perp=T_0$. For uniform equilibrium density of ions, the
self-consistent equilibrium \emph{trapped} electron distribution
is \citep{Hutchinson2019a} (for negative $W_{e\parallel}$)
\begin{equation}
  \label{eq:f0sech4}
  f_{t\parallel 0}=
         f_{e0}\left[{2\over\sqrt{\pi}}\sqrt{-W_{e\parallel}}
         +{15\over16}\sqrt{\pi\over\psi}W_{e\parallel}
         +\exp(-W_{e\parallel}){\rm erfc}(\sqrt{-W_{e\parallel}})\right],
\end{equation}
where $f_{e0}$ is the untrapped electron distribution value at zero
energy. It will be shown later that ion equilibrium density
nonuniformity gives an adjustment to the trapped electrons, but it is
fairly small at relevant $\psi$ for threshold ion distribution shapes.

The background ion parallel velocity distribution is taken as the sum
of two Maxwellian components of temperature $T_i$ shifted by
$\pm v_s$ and $v_s$ will be stated in units of
$\sqrt{T_i/m_i}$. Since ions are repelled by the hole, at some
energies one must account for reflection of ions; and the contribution
to force will then be different on the two sides of the hole if the
ion distribution is asymmetric. Moreover an equilibrium will exist at
only one hole velocity relative to the ion distribution \citep{Hutchinson2021c,Hutchinson2021d}. To avoid
having to find the equilibrium, here only symmetric ion distributions
$f_i(v)$ will be considered, for which the equilibrium hole velocity
is $v_h=0$ and the contributions from the two
sides (two incoming ion velocities) to the perturbed $\tf_i$ are
equal.

The numerical calculation proceeds as follows. For a particular $\omega$, the
required maximum number of cyclotron harmonics is found, based on
$\zeta_t$; then for each cyclotron harmonic $m$ the integrals over $z$ and
equivalently $\tau$ of eq.\ (\ref{eq:phim}), and then
(\ref{electronforce}) or (\ref{ionforce}) are carried out numerically
for fixed $W_\parallel$ (and $dW_\parallel=v_zdv_z$ so that the
phase-space element $dv_zdz\to dW_\parallel d\tau$). It proves more
accurate to integrate eq.\ (\ref{eq:phim}) as it stands, rather than
first analytically integrating by parts as was previously done \citep{Hutchinson2018a}. This improvement is especially important for the ion calculation,
in which the frequency of interest is high relative to the bounce and
transit frequencies.  After multiplying $v_zdv_z\int \Phi_m dz$ by the
terms $\omega_{m}{\partial f_\parallel\over\partial W_\parallel}$ and
$(\omega_m-\omega)f_\parallel$, the resulting $dF$, when divided by
$dv_z$ gives the differential contribution to force per unit $v_z$
which is denoted ${dF\over dv}$.

\begin{figure}\center
  (a)\hskip-1.2em\includegraphics[width=0.5\hsize]{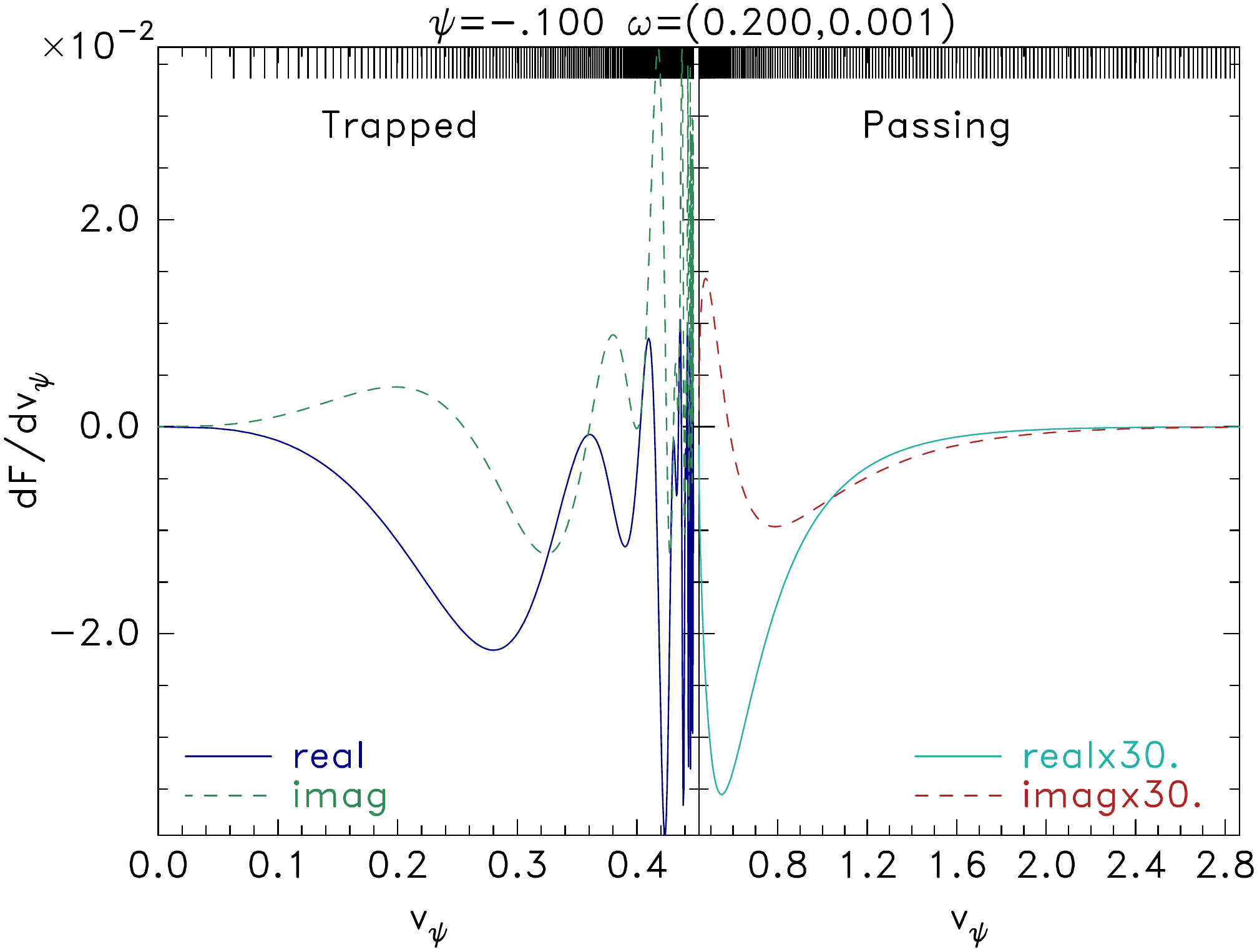}
  \ (b)\hskip-2em\includegraphics[width=0.5\hsize]{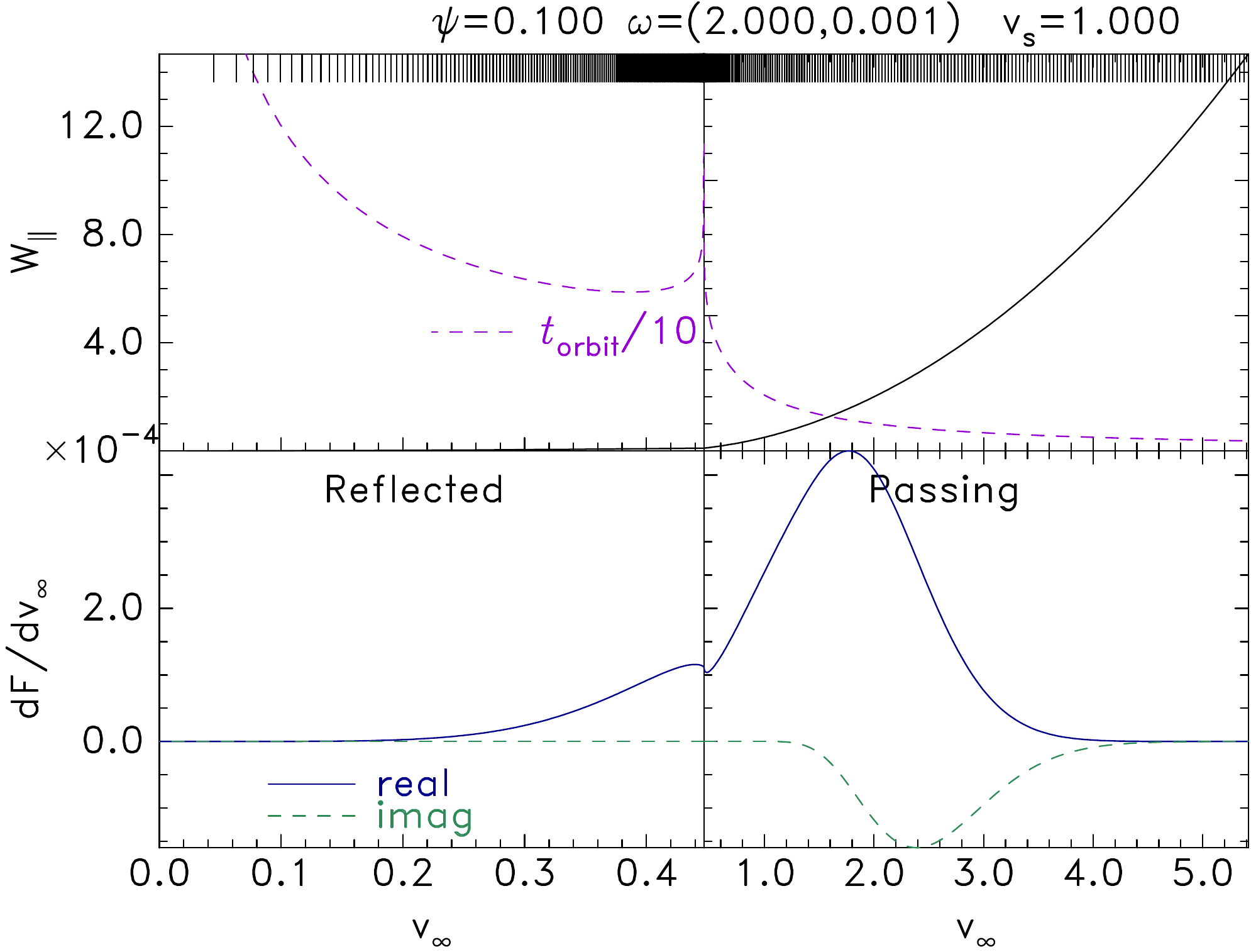}
  \caption{\label{FWplots}Illustration of the contributions to force
    $dF/dv$ as a function of parallel velocity. (a) For the attracted
    species, as a function of $v_\psi$ the velocity at the hole center
    $\phi=\psi$. Orbits are trapped for $v_\psi<\sqrt{2|\psi|}$. Note
    that passing force has been multiplied by 30 for visibility; so it
    is small. And its velocity axis scale is also different. (b) For
    the repelled species as a function of the distant velocity
    $v_\infty$, in which the upper panels show the energy and the
    orbit duration $t_{orbit}$. The non-uniform velocity integration
    mesh used is indicated at the top of each plot.}
\end{figure}
In Fig.\ \ref{FWplots} are shown illustrative cases of ${dF\over dv}$
as a function of $v_z$, for $k=0$ ($m=0$). We observe that the
\emph{attracted} species force (a) is dominated by the trapped
particles, the (single Maxwellian) passing contribution being
practically negligible. Near the trapped-passing boundary
($W_\parallel=0$ where $v_\psi=\sqrt{2\psi}$) oscillations require
fine mesh spacing to resolve, but their contribution is not strong
because of the limited extent in $v_\psi$ and significant
cancellation. By contrast, the \emph{repelled} species (b), normally
ions, (whose plotted parameters are normalized to the ion thermal
speed and plasma frequency) for which higher frequencies are dominant,
has no such oscillations. However, its resolution challenges lie in
the oscillations during integration of $\Phi_m$ on the spatial mesh,
which are not visible in this plot. For this repelled species the time
duration $t_{orbit}$ of the orbit (to travel from and back to the
chosen integration boundary) has a cusp at marginal reflection, and
becomes very long for reflected orbits at lower energy, which dwell
near reflection. The resulting averaging over fast $\phi_1$
time-oscillations rapidly suppresses the perturbed force; and so
passing orbit contributions dominate.

\begin{figure}[ht]
  \center
  \includegraphics[width=0.5\hsize]{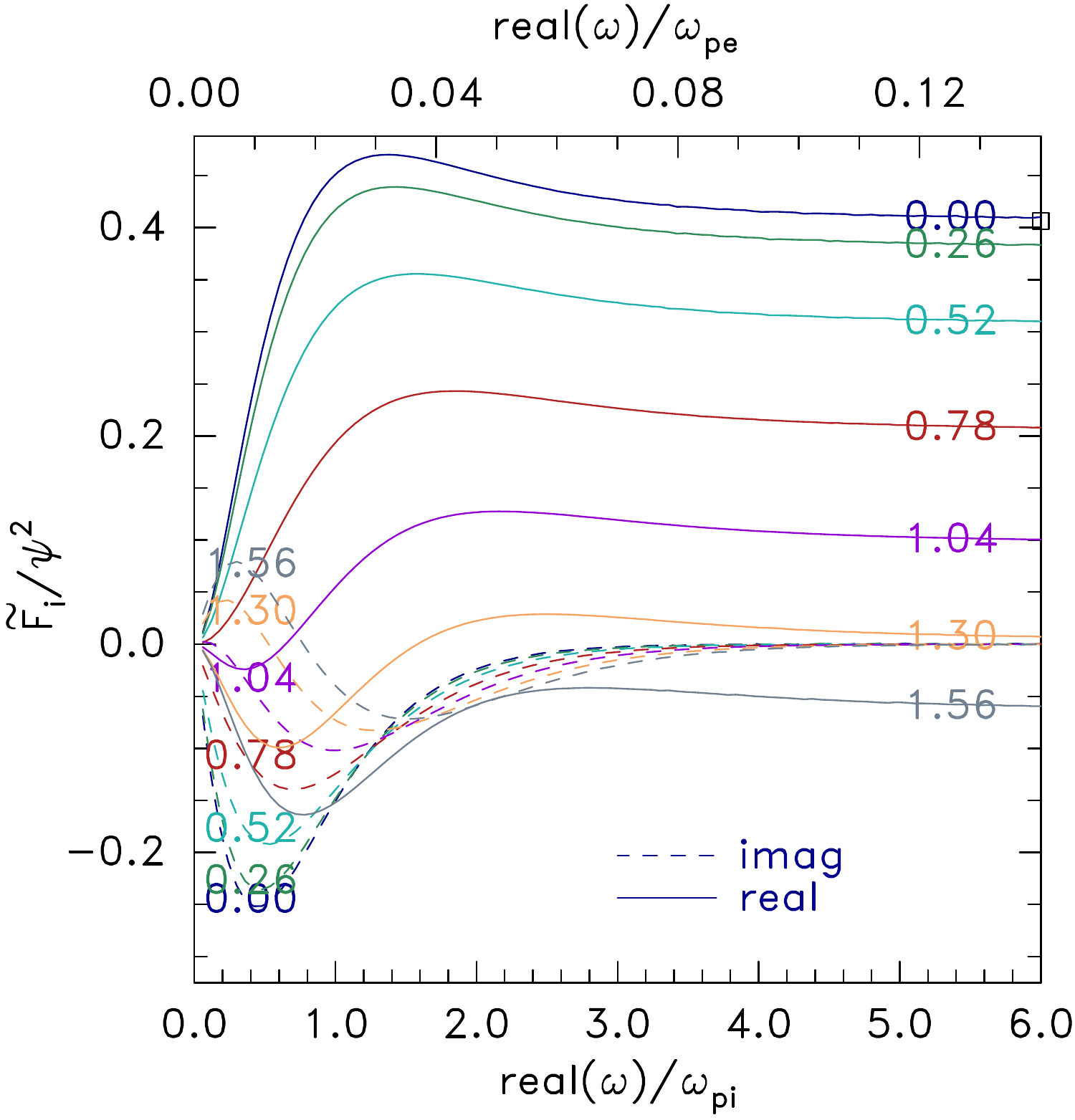}
  \caption{\label{fiofomega}The real and imaginary parts of the ion
    force as a function of real $\omega$ for several different values
    of $v_s$ labelling the curves. At high frequency, the force
    for an effectively immobile Maxwellian distribution $v_s=0$
    is shown by the square point, and agrees with the corresponding
    numerical integration line. The plot is independent of the value
    of hole amplitude $\psi$ when it is small (0.001 here).}
\end{figure}

To get the total force for the entire velocity distribution,
${dF\over dv}$ is multiplied by the $I_m(\zeta_t)$, and summed over
relevant harmonics $m$. This process is mathematically equivalent to
integrating over perpendicular distribution for low magnetic field,
but for $\zeta_t\to 0$ requires only $m=0$, the perpendicular
Maxwellian integration having been transferred to the Bessel
function. In this section and the next we show results only for this
limit (effectively $k=0$). We arrive at the total particle force $\tF$
exerted by the potential, at this chosen $\omega$.  The complex ion force
$\tF_i$ as a function of real frequency (strictly for small positive
imaginary part of $\omega$) is plotted in Fig.\ \ref{fiofomega},
normalized to $\psi^2$ the hole potential amplitude squared, which is
the dominant scaling.
The tendency to a real constant value at high frequency is because the
ion charge density remains stationary when the potential shifts faster
than the ions can respond ($\omega\gg\omega_{pi}$). The reversal of
the value with ion distribution shape ($v_s$) is because
$f_{i\parallel}(v_z)$ develops a local minimum at $v=0$ when the
velocity shift of the components is more than $\sim 1.1$. This is
approximately the shape at which exponential instability ceases. A way
to understand the force reversal intuitively is that the ion density
in equilibrium becomes \emph{larger} in the positive potential hole
than outside it, rather that being smaller for a low shift
(e.g. Maxwellian, $v_s=0$). A positive equilibrium ion density
deviation is repelled by the shifted potential, and the result tends
to oppose the shift of the hole (recognizing that the hole has
effectively negative inertia \citep{Hutchinson2019a}). The algebra
quantifies this intuition.

In Fig.\ \ref{fiofomega}, a square at the right hand side of the plot
represents the analytic value $F_i=(128/315)\psi^2$ of the ion force
for a small amplitude $\sech^4$ shape hole and Maxwellian ions
immobile on the timescale of the shift perturbation. Its agreement
with the numerical calculation at $v_s=0$ is one of the verification
tests of the code. Another is that independent numerical calculation
shows (reference  \citep{Hutchinson2021c} figure 2) that the
immobile-ion force is zero for small $\psi$ at $v_s=1.3$, which is
also in agreement with Fig.\ \ref{fiofomega}. Verification of the
passing and trapped force code for the attracted species consists of a
detailed quantitative comparison of the new code's force results with
equivalent calculations from the independent prior code, whose results
have been compared with analytic estimates \citep{Hutchinson2019a} and
PIC calculations \citep{Hutchinson2019}, as well as agreement with the
analytic electron force at very high frequency where even electrons
are immobile
($\tF_e\to
-({128\over35}-{512\over63}+{3200\over693})\psi^2=-0.14776\psi^2$). There
is an important adjustment in the electron force that arises,
particularly for peaked ion distributions ($v_s\lesssim 0.5$). It is
that the ion depletion contribution to the \emph{equilibrium} charge
changes the required equilibrium trapped electron depletion and thus
changes $\partial f_{e\parallel0}\over\partial W_{e\parallel}$ for
fixed $\phi(z)$. The effect is conveniently accounted for by scaling
the electron depletion, i.e.\ the difference between the trapped
distribution and a flat distribution $f_{e\parallel0}-f_{e0}$ (eq.\
\ref{eq:f0sech4}), by a single factor we shall write $1+c_i$, where
$c_i$ is proportional to the ion equilibrium density non-uniformity,
and is independent of $W_\parallel$. A fit to the ion response
previously found \citep{Hutchinson2021c} provides the value of
$c_i$\footnote{The ad hoc fitting expression used is
  $c_i=
  [1-(v_s/v_x)^\alpha]/[1+\psi/4+v_s\{v_s/[v_x+(3.3/v_s)^{1.5}]\}^\alpha]$,
  with $\alpha=1.4$ and $v_x=1.3+0.2\psi$ }. The factor $1+c_i$
approximately doubles $f_{e\parallel0}-f_{e0}$ at zero $v_s$, passes
through unity near $v_s=1.2$, and remains near 0.8 for higher
$v_s$. This approximation of the ion equilibrium effects introduces
uncertainties that are no greater than those implied by the adoption
of the $\sech^4$ potential form in the first place, and by the
assumption that the eigenmode is the shiftmode.

The
force-balance dispersion relation,
\begin{equation}
  \label{eq:dispersion}
  (\tF\equiv)\tF_e+\tF_i=F_E,
\end{equation}
which is a complex equation requiring both real and imaginary parts to
be zero, is not satisfied by an arbitrary complex $\omega$. The
$\omega$ that \emph{does} satisfy it (if any), can be found either by
contouring in the complex $\omega$ plane the real and imaginary parts
of $\tF-F_E$ and finding where their zero contours cross, or by
two-dimensional Newton iteration to find the complex root of $\tF-F_E$
(constrained by requiring $\Real(\omega)\ge 0$ and
$\Imag(\omega)\ge 0$).
\begin{figure}[ht]
(a)\hskip-1em \includegraphics[width=.49\hsize]{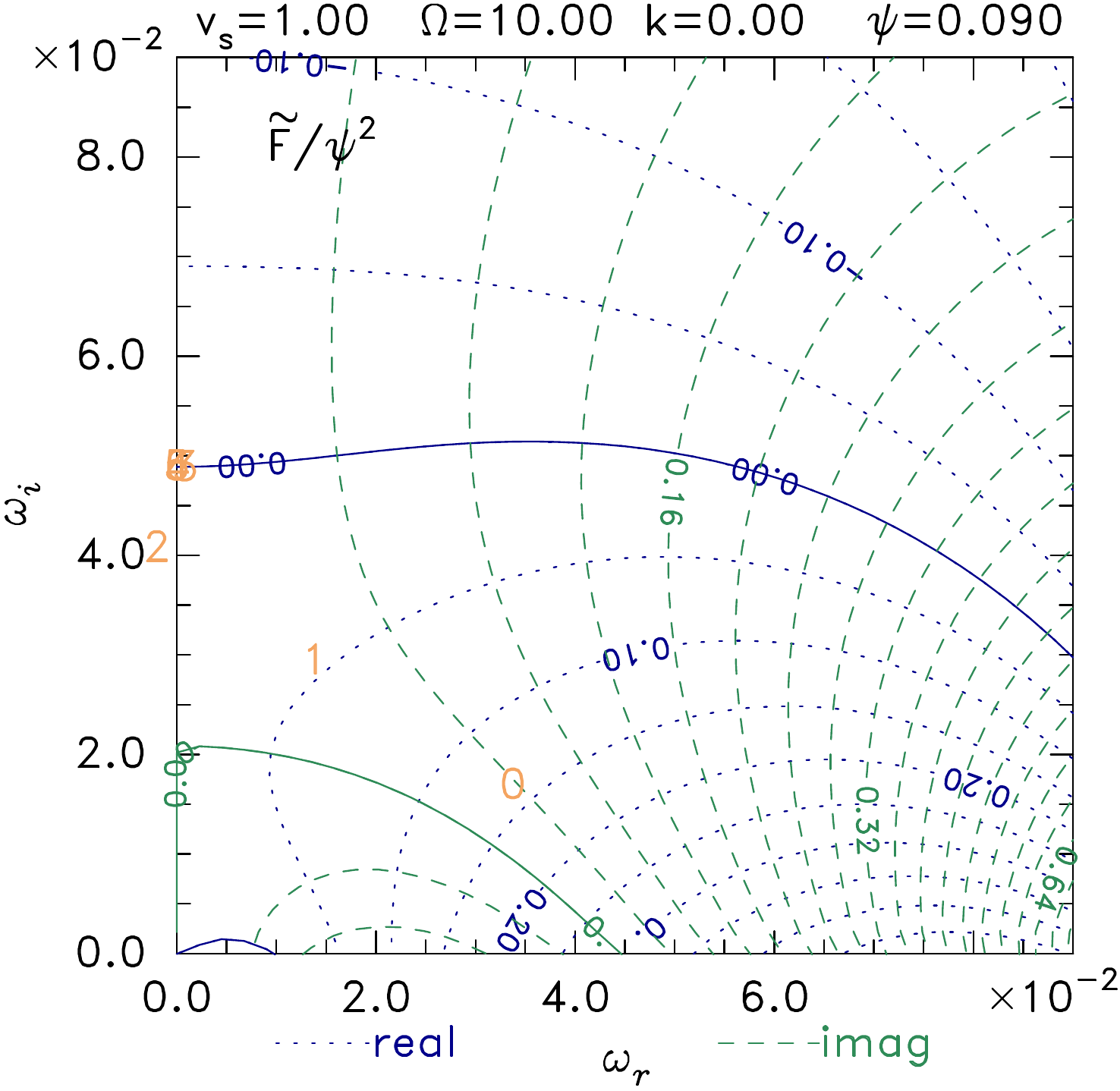}
(b)\hskip-1.5em\includegraphics[width=.49\hsize]{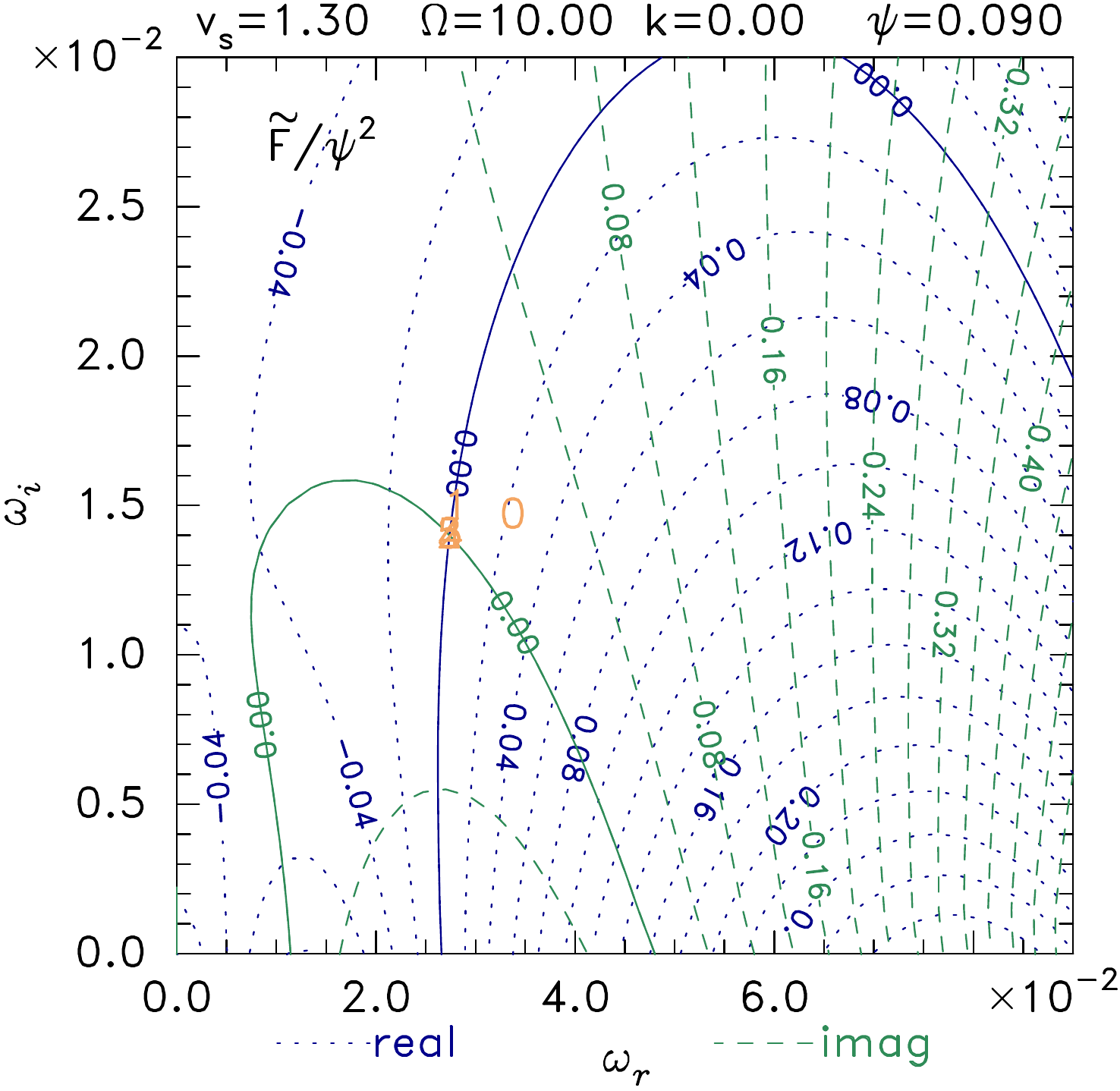}
\caption{\label{Fcont109}Contours of real and imaginary parts of total
  force $\tF$ over the relevant complex $\omega$ region. Solid
  contours have value zero. The sequence of (orange) numbers on the
  plane show where Newton iterations (starting at 0) take the
  iterative solution. It converges at the intersection of the zero
  contours, where $\tF=0$. In (a) $v_s=1$ the solution lies on the
  $\Real(\omega)=0$ boundary, where $\Imag(\tF)=0$. In (b) with larger
  ion component separation $v_s=1.3$, there is no solution with
  $\Real(\omega)=0$, but a slower growing oscillation ocurs at finite
  $\Real(\omega)$.}
\end{figure}

\section{One-dimensional Stability: Mechanism and Results}\label{sec3}

In this section we presume $k=0$, which means $F_E=0$ and motion is
effectively one-dimensional.  Figure \ref{Fcont109} shows example
contour plots of the total complex force in the complex $\omega$
plane. The complex root of the dispersion relation is where $\tF=0$,
at the intersection of the zero contours of its real and imaginary
parts.  At $v_s=1$ (a), where there is no local minimum in the ion
velocity distribution, a purely exponentially growing instability
occurs at $\omega\simeq 0.05i$. At $v_s=1.3$ (b), where a local
minimum is present, the root is a slower-growing oscillation
$\omega\simeq 0.028+0.014i$. This second type of solution was
unaddressed in the prior slow hole stability analysis.

The disappearance of the purely growing mode can be understood by
plotting the ion and electron forces along the $\Real(\omega)=0$
boundary.\begin{figure}\center
  \includegraphics[width=0.5\hsize]{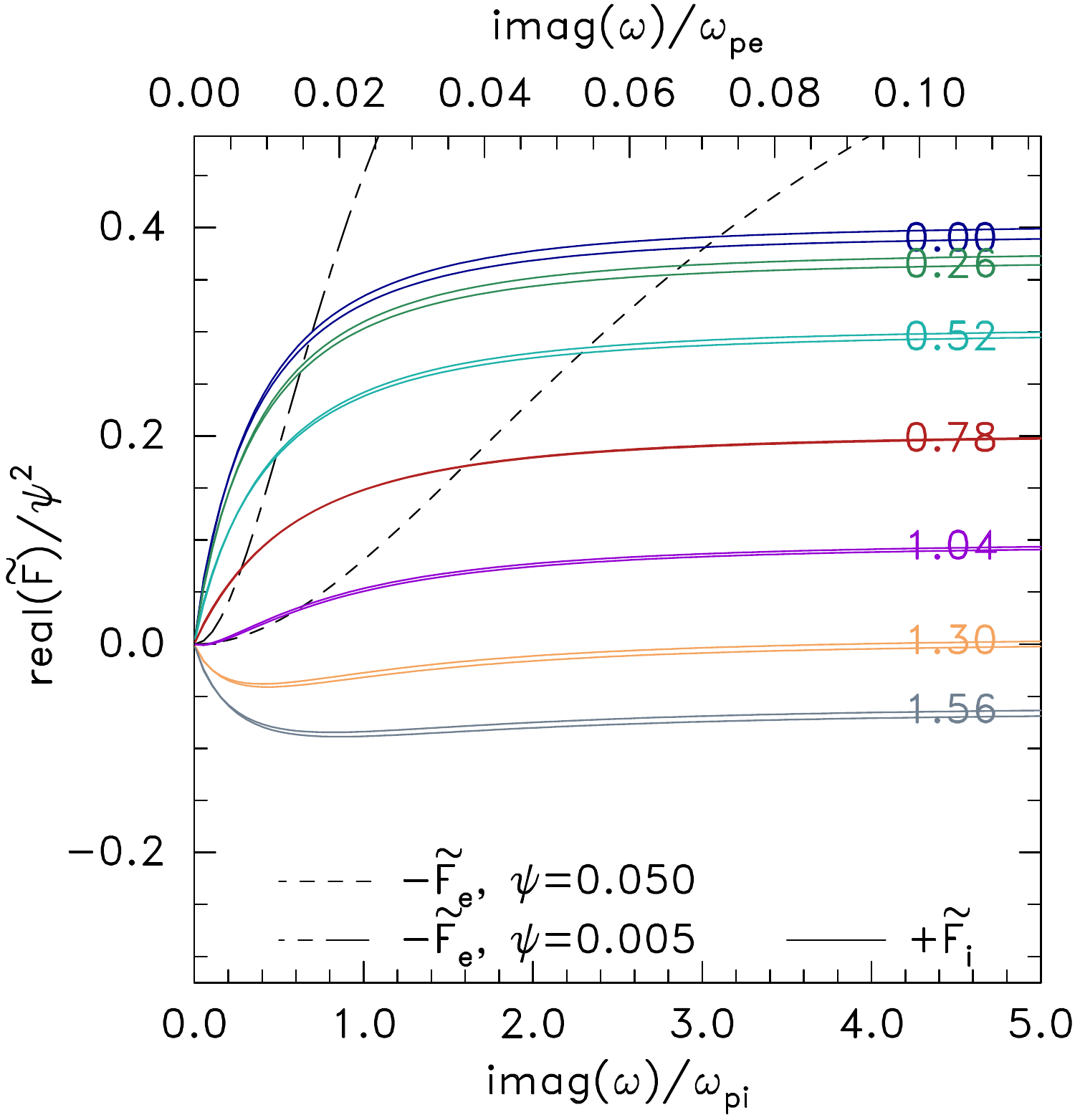}
  \caption{\label{Fiofimagomega}Scaled force $\tF/\psi^2$ as a
    function of purely imaginary $\omega$. The ion force $\tF_i$ is
    almost independent of $\psi$ but depends on $v_s$ shown by labels
    on the solid lines. The dashed curves give minus the electron
    force for the different $\psi$ values. A growing mode requires
    $\tF_i=-\tF_e$: an intersection. When $v_s$ is large enough, there
    are no intersections.}
\end{figure}
As shown in Fig.\ \ref{Fiofimagomega}, although the ion force
normalized to $\psi^2$ is very insensitive to the $\psi$-value, that
is not true of the electron force. The reason is that $F_e$ mostly
comes from trapped particles whose bounce frequency is
$\omega_b\propto \sqrt{\psi}$, and $\omega_b/\omega$ determines the
force variation. By contrast, the ion force is predominantly passing
(no ion trapping), and has much weaker scaled dependence on $\psi$.
When $v_s\gtrsim1$ no intersection ($\tF_i=-\tF_e$) occurs for small
enough $\psi$, and there is no instability having $\Real(\omega)=0$.
All this is consistent with the previously published slow hole
stability analysis concentrating on the high and negligible real
frequency regions of the $\omega$ plane, which concluded that a local
minimum in $f_i(v)$ was necessary for stability.

However, the \emph{oscillatory} solution of Fig.\ \ref{Fcont109}(b) can
remain even when there is a local $f_i$ minimum if the
$\Imag(\tF)=0$ contour lies above the real $\omega$-axis.
\begin{figure}
\center  \includegraphics[width=.5\hsize]{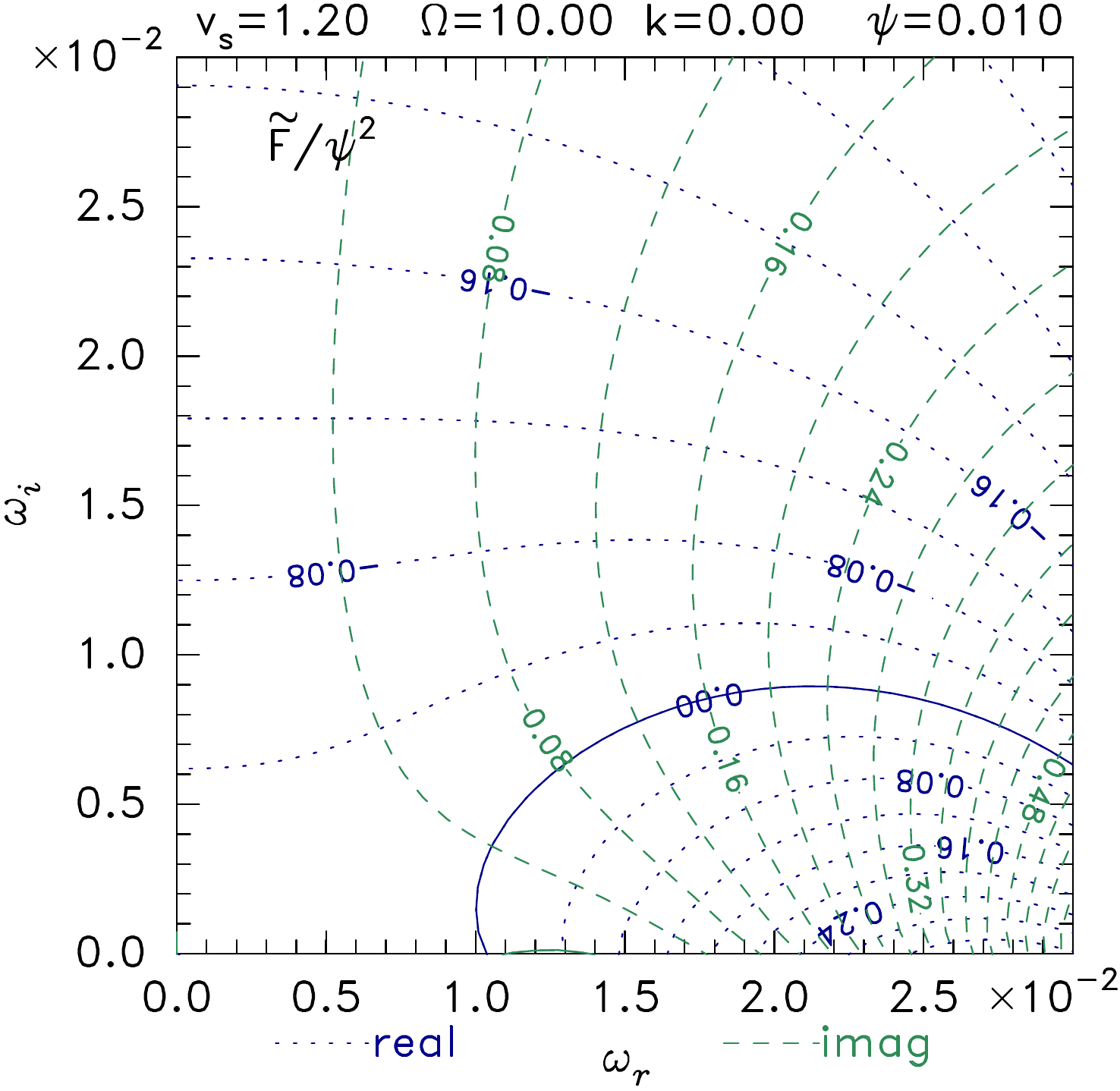}
  \caption{\label{Fcont1201}Force contours for a stable case at small
    $\psi$ with sufficient ion component shift to give a local
    minimum. The negative imaginary contour is disappearing below the
    real axis removing the root of $\tF$ from the unstable
    domain.}
\end{figure}
The solution disappears if the hole amplitude is small enough, because
the $\Imag(\tF)=0$ contour disappears below the real axis, as Fig.\
\ref{Fcont1201} illustrates. The root itself, where $\Real(\tF)=0$ is
zero as well as $\Imag(\tF)$, likewise disappears. The case shown has
just sufficiently low $\psi$ for total stability.  Newton iterations
bump up against the $\Imag(\omega)\ge 0$ constraint without finding a
zero. Decreasing $\psi$ or increasing $v_s$ moves the (barely
visible) imaginary contour even further down out of the unstable
domain.

Thus, there are really two requirements for the slow electron hole to
be fully stable, to oscillatory as well as purely growing modes: (A)
that there be a local minimum in the ion distribution $f_i(v)$
($v_s\gtrsim 1.1 v_{ti}$), and (B) that the hole's potential amplitude
$\psi$ should be small enough ($\psi\lesssim 0.01 T_e/e$).

\begin{figure}
  \center
  \includegraphics[width=0.6\hsize]{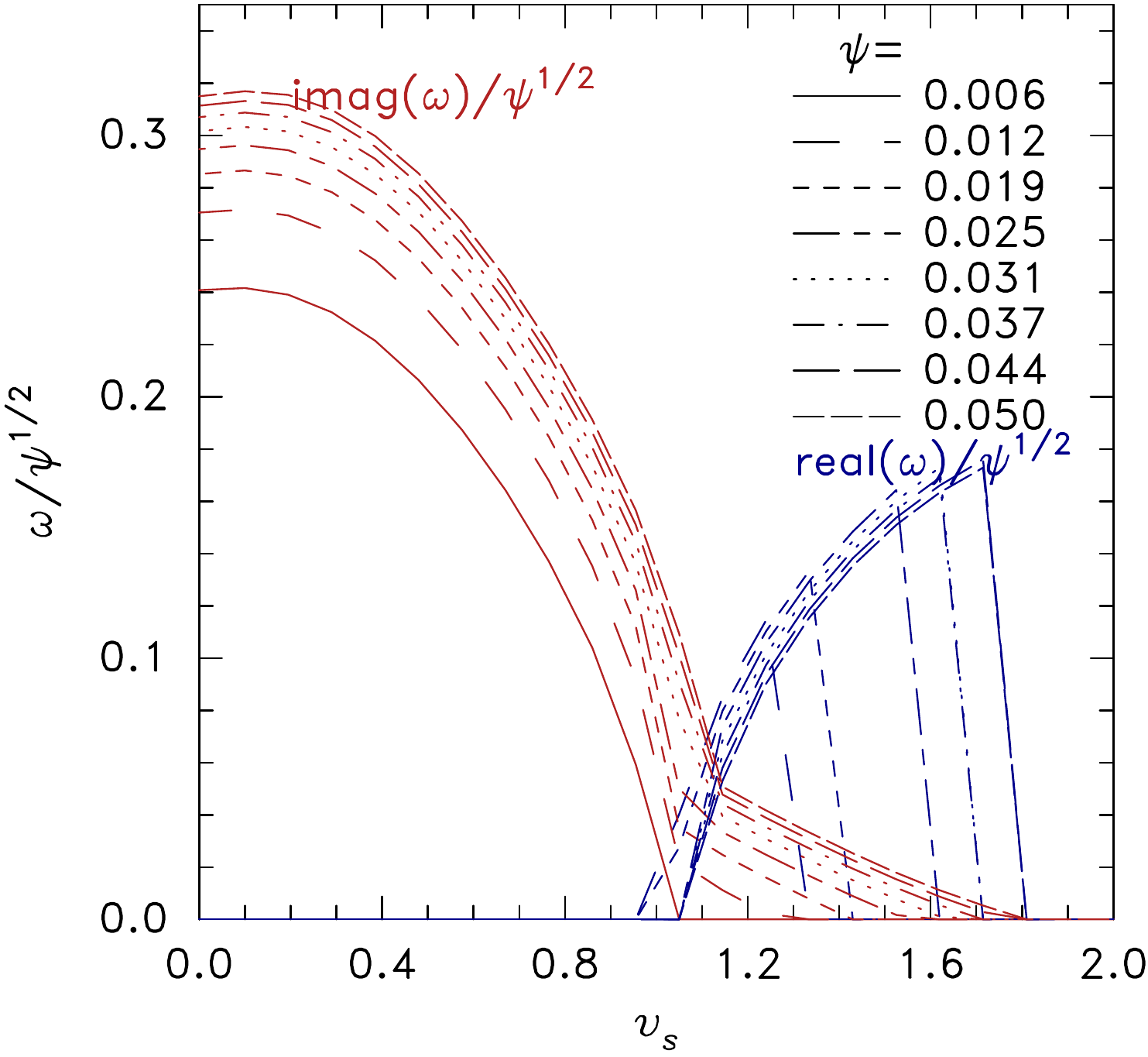}
  \caption{\label{omegaofv1}The real and imaginary parts of unstable
    mode frequency versus ion distribution shape ($v_s$), for a range of peak
    potential amplitudes ($\psi$).}
\end{figure}

More quantitatively, a range of results is summarized in Fig.\
\ref{omegaofv1}.
At low $v_s$, the shift mode is purely growing, with a scaled rate
$\Imag(\omega)\simeq 0.3\psi^{1/2}$ for a Maxwellian ($v_s=0$) ion
distribution. The transition to oscillation occurs close to to
$v_s=1$, and the growth rate $\Imag(\omega)$ decreases quickly (as the
local minimum of $f_i(v)$ develops) by a total factor of approximately
10. As that minimum deepens ($v_s$ increasing further), the growth
rate drops more slowly and the oscillation frequency $\Real(\omega)$
increases until at a depth dependent on $\psi$ the growth becomes zero
and the instability disappears. If the amplitude is small enough
$\psi\lesssim 0.01$, there is no oscillatory instability,
and developing a local bare minimum in $f_i$ is sufficient for stability.

\section{Variation with other parameters}\label{sec4}

\subsection{Ion Temperature}

When ion temperature $T_i$ is different from electron temperature
($=T_0$), the ion force is changed. Referring to equations
\ref{ionforce} and \ref{eq:fimagnetic}, the quantities that are
directly affected are
${\partial f_{i\parallel 0}\over dW_{i\parallel}}$,
${f_{i\parallel0}\over T_{i\perp}}$, and $\zeta_{it}$. In this
subsection we will not address $\zeta_{it}$ changes with $T_i$. They
give negligible effect in the strong ($\zeta_t\to 0$) and weak
($\zeta_t>20$) magnetic field limits. For simplicity supposing the two
parallel-shifted Maxwellian ion components remain isotropic, the
effect of $T_i$ changes on
${\partial f_{i\parallel 0}\over dW_{i\parallel}}$ and
${f_{i\parallel0}\over T_{i\perp}}$ is to divide the ion force by
$T_{i}/T_0$; because the changes arising from $f_{i\parallel0}$ itself
(reducing its height $\propto 1/\sqrt{T_i}$, and increasing its
velocity width $\propto \sqrt{T_i}$), cancel each other out. Thus, the
ion density perturbations, both in equilibrium and resulting from
$\phi_1$, are multiplied by $T_0/T_i$.  And since we are citing $v_s$
in ion thermal units, the ion velocity distribution relative
\emph{shape} is independent of $T_i$. Therefore ion temperature can be
accounted for by multiplying the ion force $\tF_i$ and the equilibrium
ion density non-uniformity by $1/T_i$.

Because the equilibrium ion density is changed by $T_i$, the electron
force is also changed indirectly by adjustment of the required
electron trapped distribution for equilibrium. The trapped electron
equilibrium deficit is proportional to $(1+c_i/T_i)$ and the electron
force is scaled by the same factor, which varies more slowly than the
ion force factor $1/T_i$.

Higher ion temperature therefore decreases the force on the ions
relative to that on the electrons. Referring back to Fig.\
\ref{Fiofimagomega}, the different scaling of these curves has the
effect of moving the intersection $\tF_i=-\tF_e$ to lower imaginary
values of frequency, that is of reducing the growth rate, as $T_i$
increases.

\begin{figure}
  \center
  \includegraphics[width=0.6\hsize]{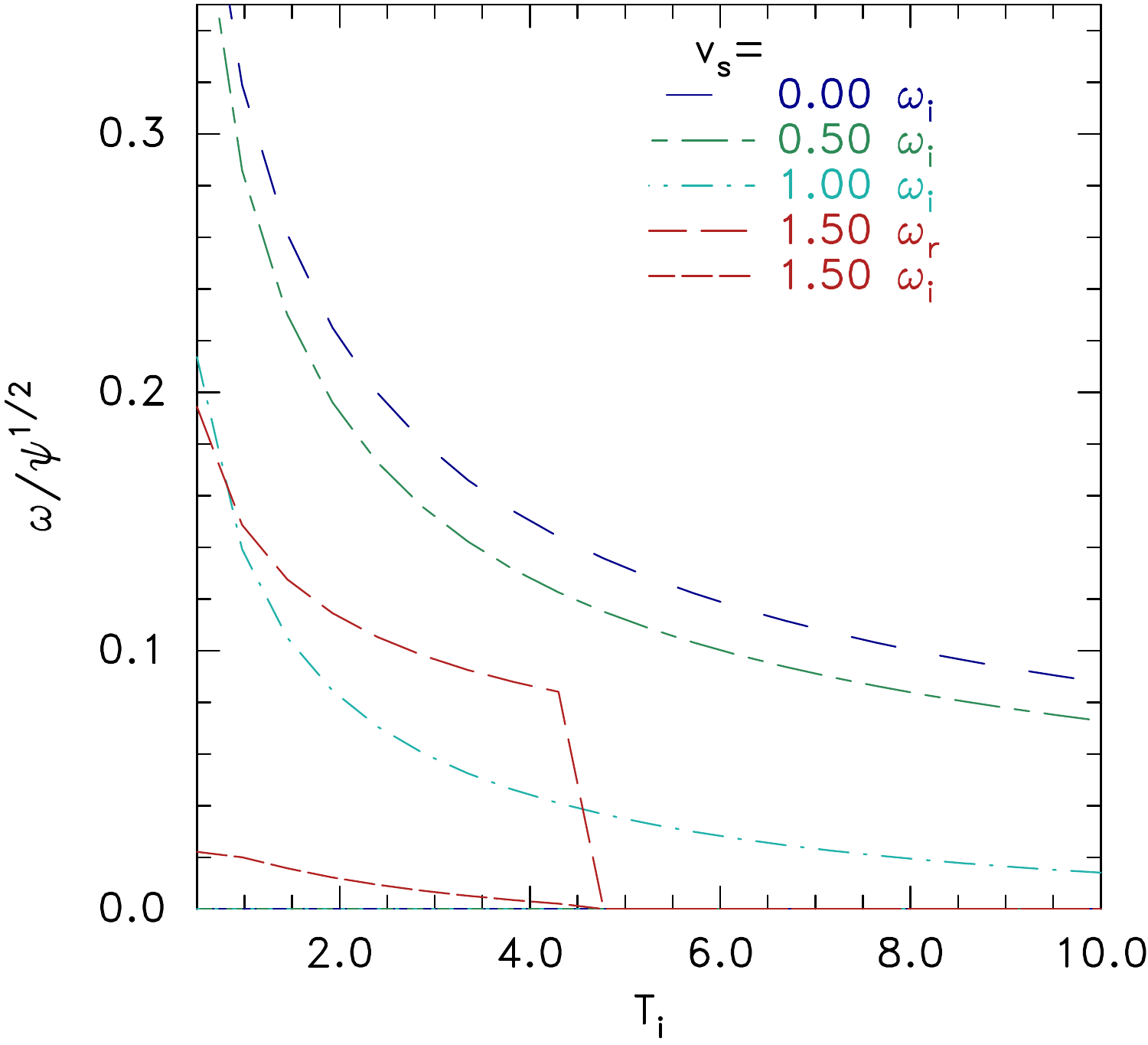}
  \caption{\label{omegaofTi}Variation of unstable frequencies
    ($\omega_i$ imaginary, $\omega_r$ real if non-zero) with ion
  temperature, for different $v_s$. $\psi=0.05$.}
\end{figure}

Figure \ref{omegaofTi} illustrates the resulting variation of the
instability frequency found by full numerical integration for
$k=0$. Increasing $T_i$ always decreases the growth rate.  For
$v_s\lesssim 1.1$ the real frequency is zero, and is not plotted. For
the oscillatory instability regime, $v_s\gtrsim 1.1$, increasing $T_i$
can lead to a full stabilization: when $\psi=0.05$ and $v_s=1.5$ at
$T_i\simeq 5$.

\subsection{Transverse variation: Multidimensional Instability}

Multidimensional instability here means situations in which although
the equilibrium is independent of transverse position, the
perturbation varies transversely: $k\not=0$.  And in this section we
shall suppose that the magnetic field is strong enough that still only
the $m=0$ cyclotron harmonic matters because $\zeta_t$ remains
small. In that case the only new effect is that the transverse Maxwell
stress force $F_E$ (eq.\ \ref{eq:forcebalance}) is non-zero and must
be included in the dispersion equation
\ref{eq:dispersion}. Again suppressing the linear $\xi$ factor, for the
shift mode, it becomes
\begin{equation}
  \label{FEeq}
  F_E= k^2\int\left(d\phi_0\over dz\right)^2 dz= k^2 {128\psi^2\over 315},
\end{equation}
where the final equality applies for the specific $\sech^4$ potential shape.
Therefore, if we consider the transverse wave-number $k$ to be an
arbitrary (real) choice, then we must examine the effect of including
an $F_E$ that is real and positive, but of arbitrary magnitude. Thus,
we must allow the real part of $\tF$ to be positive.

The contour plots of Fig.\ \ref{Fcont109} (for example) allow us
immediately to deduce what happens to the unstable root as $F_E$
increases from zero: it becomes the intersection of the unchanged
solid $\Imag(\tF)=0$ contour with a different real contour
$\Real(\tF)=F_E$, instead of zero. For example, in the case of Fig.\
\ref{Fcont109}(a) if a value $F_E=0.1\psi^2$ is included, the relevant
real contour is the one labelled 0.1. The root is then instead at
$\omega\simeq(2.4+1.5i)\times10^{-2}$, giving a slower growing
oscillatory instability rather than the purely growing one for
$k=0$. By eq.\ \ref{FEeq} $F_E/\psi^2=0.1$ corresponds to
$k=\sqrt{0.1\times315/128}=0.496$, which is at the limit of high $k$,
short wavelength; smaller $k$-values will have even less effect.

Similarly in Fig.\ \ref{Fcont109}(b) as $k$ is increased, the
relevant real contour moves to the right, and the root to higher real
and lower imaginary frequency. For the case of Fig.\ \ref{Fcont1201},
the motion of the real contour to the right can give the root a
slightly positive imaginary frequency, and so induce a very slow
instability where there was none before. But futher increase of $k$
brings the intersection past the region ($1.1<\omega_r<1.3$) where the
imaginary contour is positive, thus restabilizing it.

To summarize, finite $F_E$ arising from non-zero $k$ \emph{always}
acts in such a direction as to make purely growing instabilities more
stable, and \emph{usually} does so even for oscillatory modes. We therefore
do not spend further effort here to quantify in separate detail $F_E$-effects
because almost always the $F_E=0$ situation already treated is the
most unstable. Nevertheless $F_E$ is fully included in the results of
the following section.

\subsection{Magnetic Field Strength}

When $k\not=0$ \emph{and} $\Omega$ is not very large, the Bessel
function argument may not be small, and the sum over harmonics $m$ may
include many relevant terms. In this regime, the stability can depend
on the magnetic field strength. Since the ion thermal Larmor radius $r_L$
is greater than the electron by a factor $\sqrt{m_i/m_e}$ (for
comparable temperature), the ions will generally experience
$\zeta_t(=kr_{L})\gtrsim 1$ first. The number of harmonics making
significant contribution, when substantially greater than one, may be
estimated by noting that $\zeta_t{\rm e}^{-\zeta_t^2}I_m(\zeta_t^2)$,
regarded as a function of integer $m$, is approximately
a Gaussian ${\rm e}^{-v_m^2/2}/\sqrt{2\pi}$, where $v_m=m/\zeta_t$. We therefore
require contributions up to $v_m\sim 3$, i.e.\ $m\sim 3\zeta_t$. However
with this velocity identification,
$\omega_m=\omega+m\Omega=\omega+kv_m$, so writing $dv_m=1/\zeta_t=\Omega/k$,
eq. \ref{eq:fimagnetic} becomes
\begin{equation}
  \label{harmsum}
  \begin{split}    
  &\sum_{m=-\infty}^\infty i\left[\omega_{m}
  {\partial f_{\parallel0}\over \partial W_{\parallel}}
  +(\omega_{m}-\omega) {f_{\parallel0}\over T_{\perp}}\right]
  q\Phi_m {\rm e}^{-\zeta_{t}^2}I_m(\zeta_{t}^2)\simeq\\
  &\sum_{m=-\infty}^\infty i\left[(\omega+kv_m)
  {\partial f_{\parallel0}\over \partial W_{\parallel}}
  +kv_m {f_{\parallel0}\over T_{\perp}}\right]
  q\Phi_0(\omega+kv_m){\rm e}^{-v_m^2/2}dv_m.
  \end{split}
\end{equation}
The second form is a finite difference approximation of the
unmagnetized integral expression: $\sum_{m=-\infty}^\infty dv_m\simeq
\int_{-\infty}^\infty dv$. Thus, as physical considerations dictate,
the limit of small $\Omega/kv_t$ gives simply an unmagnetized integral over
the $k$-component of the perpendicular velocity distribution. This is
often the case for the ions.

Although the numerical routine\footnote{RIBESL from the library
  TOMS715} evaluating the higher order Bessel functions (by backward
recurrence) works well only up to approximately $I_{60}$, a maximum
value of $|m|=60$ is sufficient for the harmonic sum to represent the
integral limit very well. It is then effectively already in the zero
field limit.  Therefore the present evaluation of $\tf$ does not
bother to switch to higher-resolution finite-difference integral
representation, and instead artificially limits how small an $\Omega$
is permitted, and uses eq.\ \ref{harmsum} with an effective value of
$\Omega$ no smaller than what requires a maximum $|m|\le60$, i.e.\
$\zeta_t\lesssim 20$. (The required effective $\Omega$ lower limit is
different for electrons and ions.)

\begin{figure}[htp]
  \ (a)\hskip-1.5em\includegraphics[width=0.49\hsize]{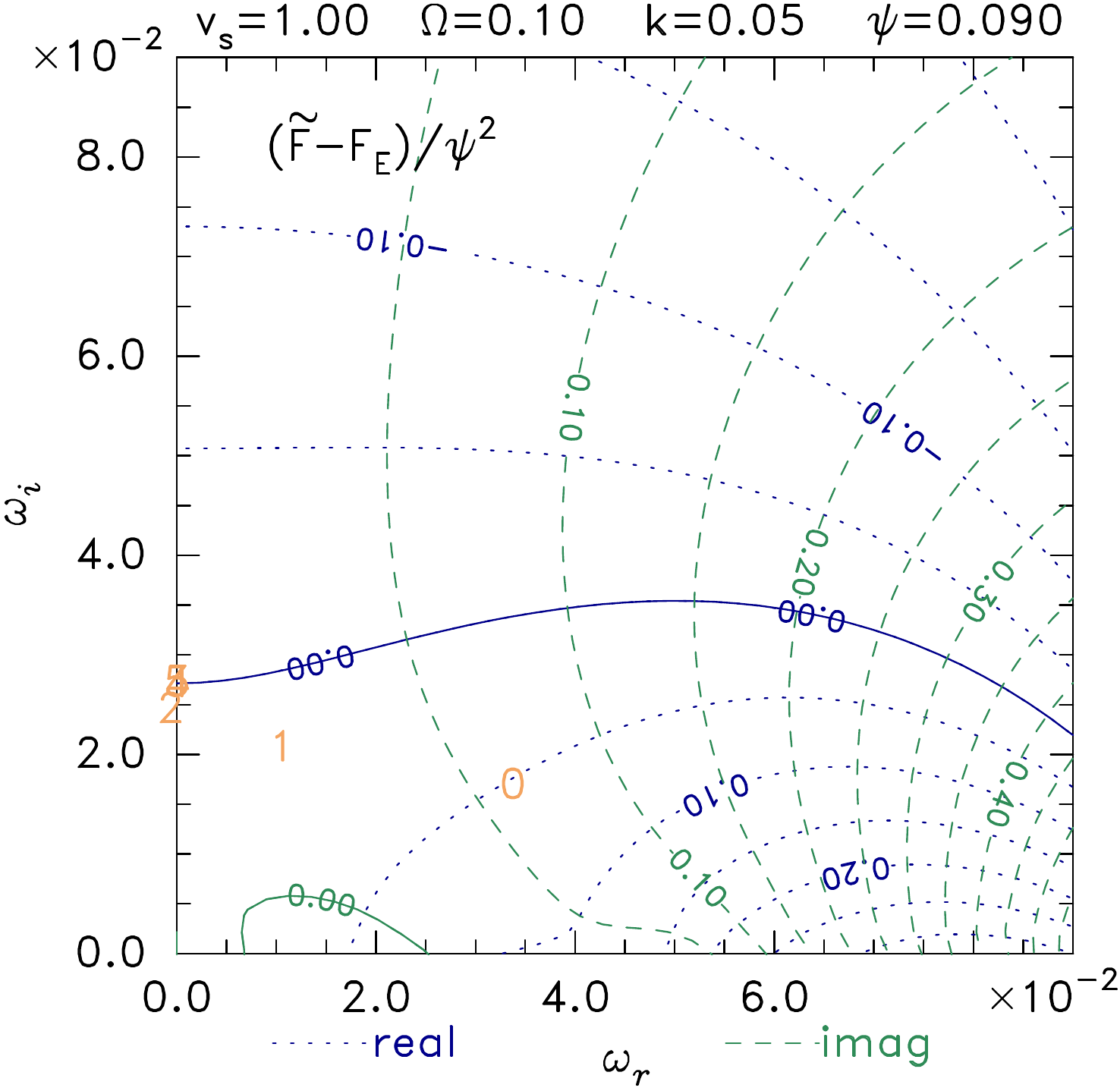}
  \ (b)\hskip-1.5em\includegraphics[width=0.49\hsize]{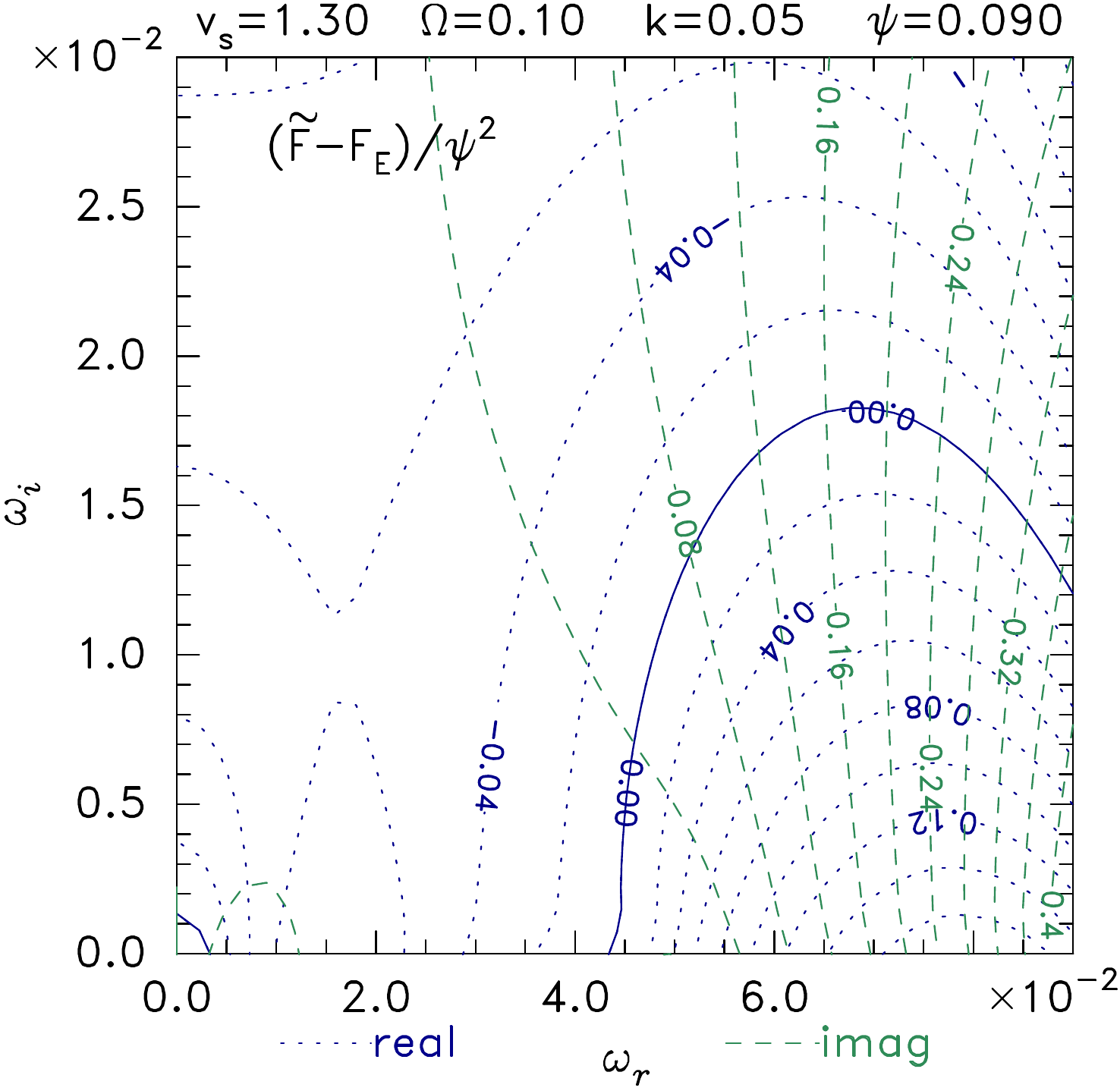}
  \ (c)\hskip-1.5em\includegraphics[width=0.49\hsize]{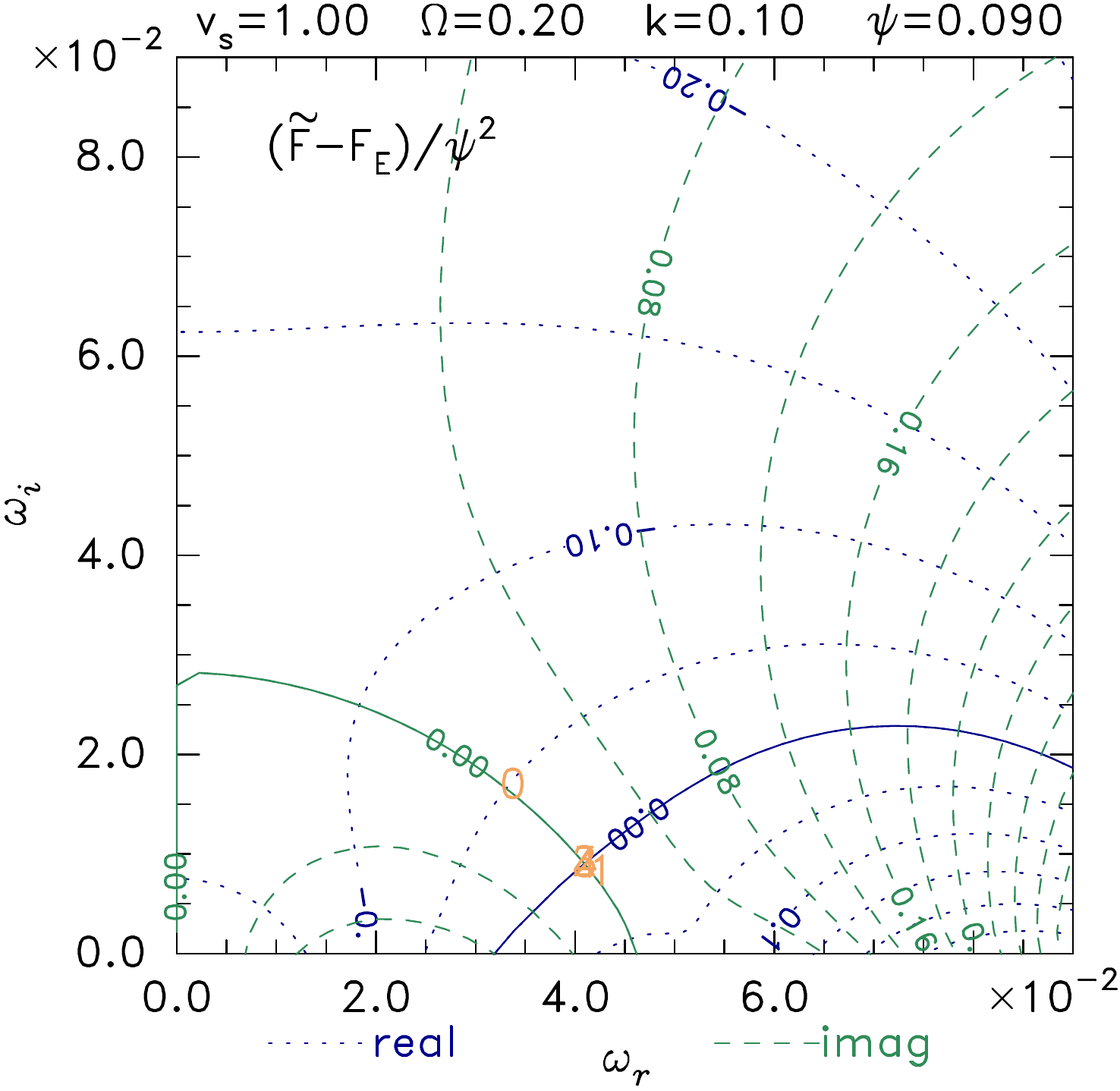}
  \ (d)\hskip-1.5em\includegraphics[width=0.49\hsize]{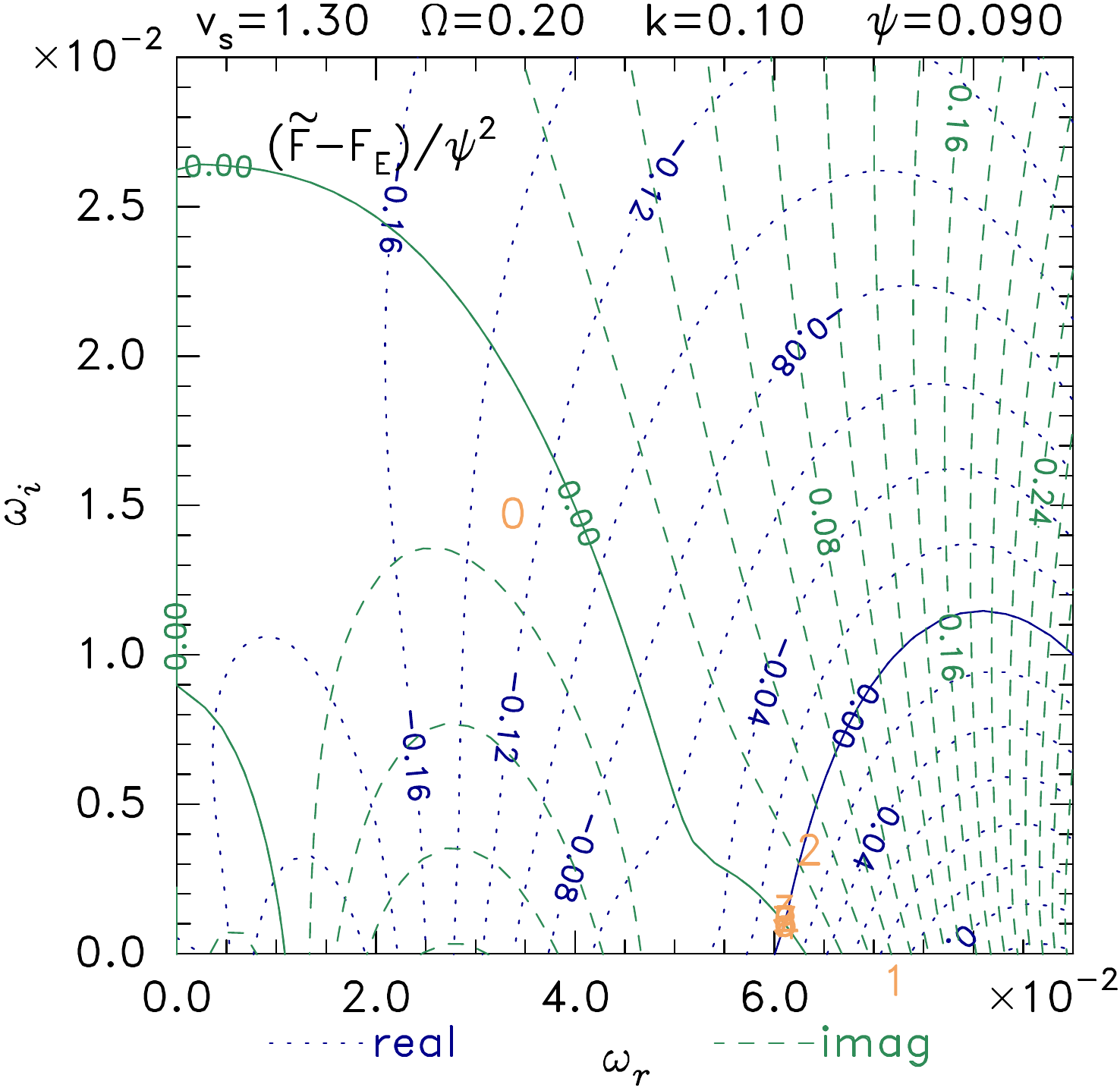}
  \caption{\label{cont109k1}Force contours with non-zero $k$ and low
    enough $\Omega$ to affect stablity.}
\end{figure}

Figure \ref{cont109k1} shows the effect of non-zero $k$ and
$\Omega=2k$ on the cases of Fig.\ \ref{Fcont109}, with which it should
be compared. Figure \ref{cont109k1}(a) and (b) have $k=0.05$ and show
mostly a movement of the contours downward toward more negative
imaginary part ($\omega_i$) of $\omega$. (a) has a relatively small
growth rate reduction; but (b), which has a local $f_i$ minimum, shows
complete stabilization of the prior unstable oscillatory mode. Figure
\ref{cont109k1}(c) and (d) have the same ratio $\Omega/k$ as (a) and
(b) but twice as large values: $k=0.1$. In (c) the mode oscillates
with real frequency $\omega_r\simeq 0.04$. In (d) there is again
tendency to higher real frequency, compared with Fig.\ \ref{Fcont109},
but a very slow growing oscillatory instability remains, unlike (b)
which was fully stabilized.  The differences between (a) and (c) and
between (b) and (d) are practically all differences in the electron
force $\tF_e$, with a modest contribution from $F_E$
differences. There is essentially negligible change in $\tF_i$,
because $\zeta_t$ is so large that we are in regime of
quasi-continuous perpendicular integration, and the value of $kv_{ti}$,
much smaller than typical $\omega$, gives little effect.

\begin{figure}
  \ (a)\hskip-1.5em\includegraphics[width=0.49\hsize]{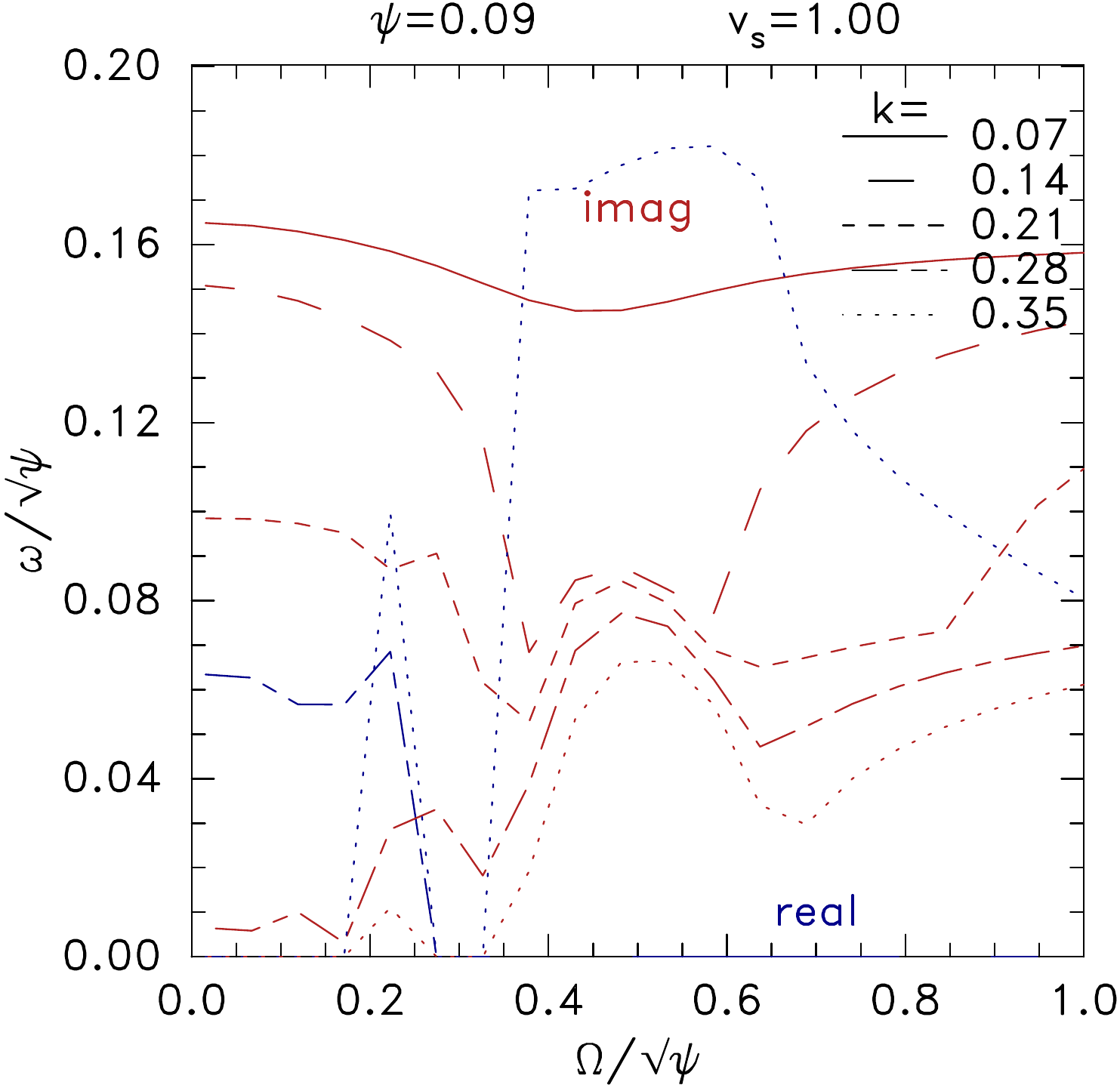}
  \ (b)\hskip-1.5em\includegraphics[width=0.49\hsize]{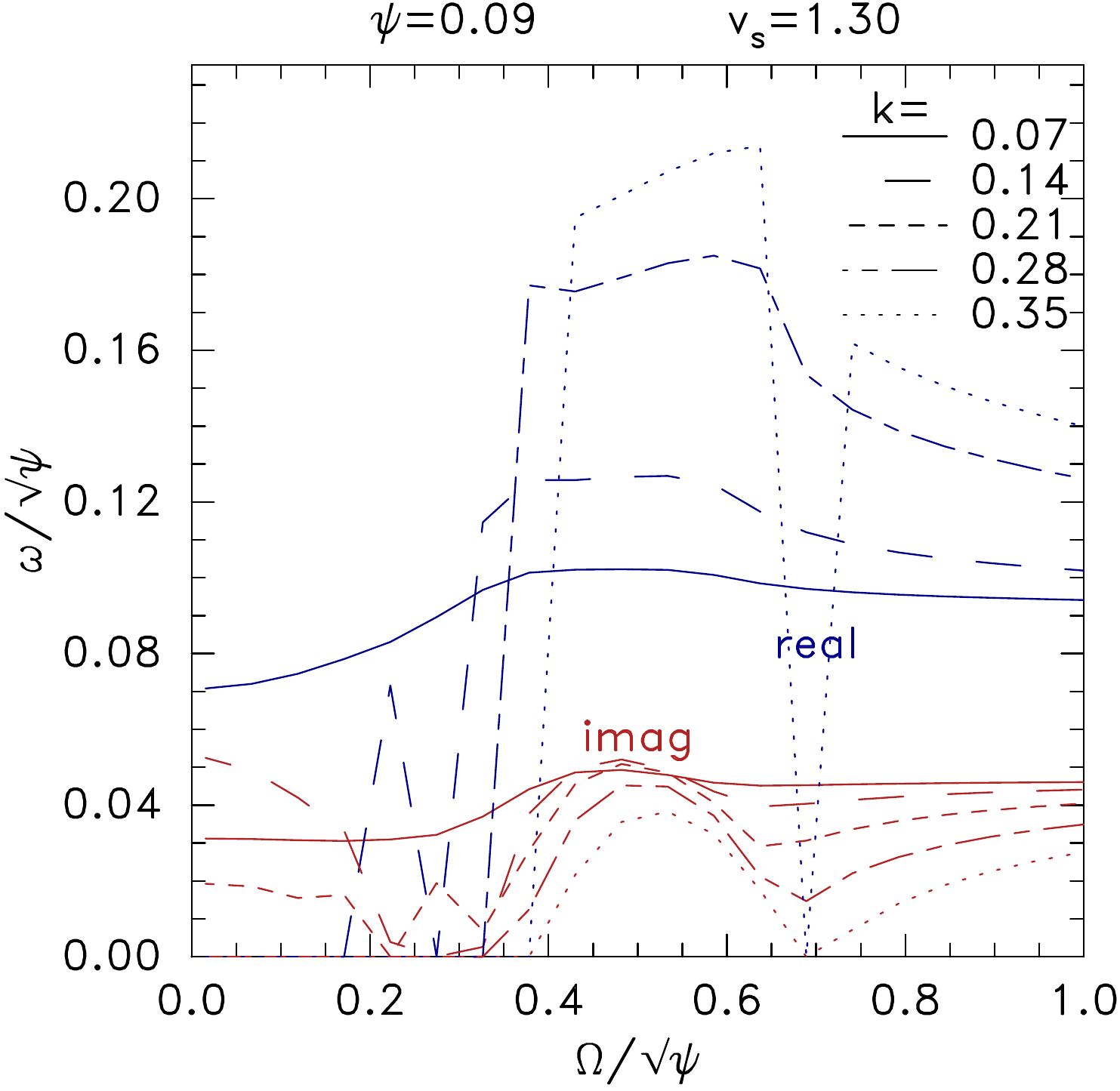}
  \caption{\label{ovo1}Dependence of the real and imaginary parts of
    the instability frequency on magnetic field strength, $\Omega$,
    for a range of transverse wavenumbers $k$. (a) Single humped
    $f_i(v)$ $v_s=1$ ; (b) double-humped $v_s=1.3$, for a moderately
    deep hole $\psi=0.09$.}
\end{figure}
Figure \ref{ovo1} shows the real and imaginary parts of the dispersion
solution $\omega$ as a function of magnetic field strength
$\Omega$. The different $k$ values control the strength of the effect
of transverse variation. The lowest $k=0.07$ value (solid lines) shows
rather weak variation from the $k=0$ case (not shown) which gives a
solution independent of $\Omega$ and equal to the high-$\Omega$ limit
of the solid lines. In (a) the single-humped ion velocity distribution gives a
purely growing mode ($\Real(\omega)=0$) at low $k$ (0.07 and below)
with growth rate approximately 0.16. All higher $k$ values have growth
rates smaller than this. Some of them make transitions to oscillatory
($\Real(\omega)\not=0$) modes in the middle of the $\Omega$ range, with
growth rates typically half as great. In (b) for a double-humped ion
distribution, similar trends are evident, but the growth rate
is considerably smaller $\lesssim 0.05$, and the real frequency never
reverts to zero at high $\Omega$, consistent with the previous
observation that for such distributions one-dimensional modes are all
oscillatory, and are unstable at this $\psi$ value. 

\begin{figure}
  \ (a)\hskip-1.5em\includegraphics[width=0.49\hsize]{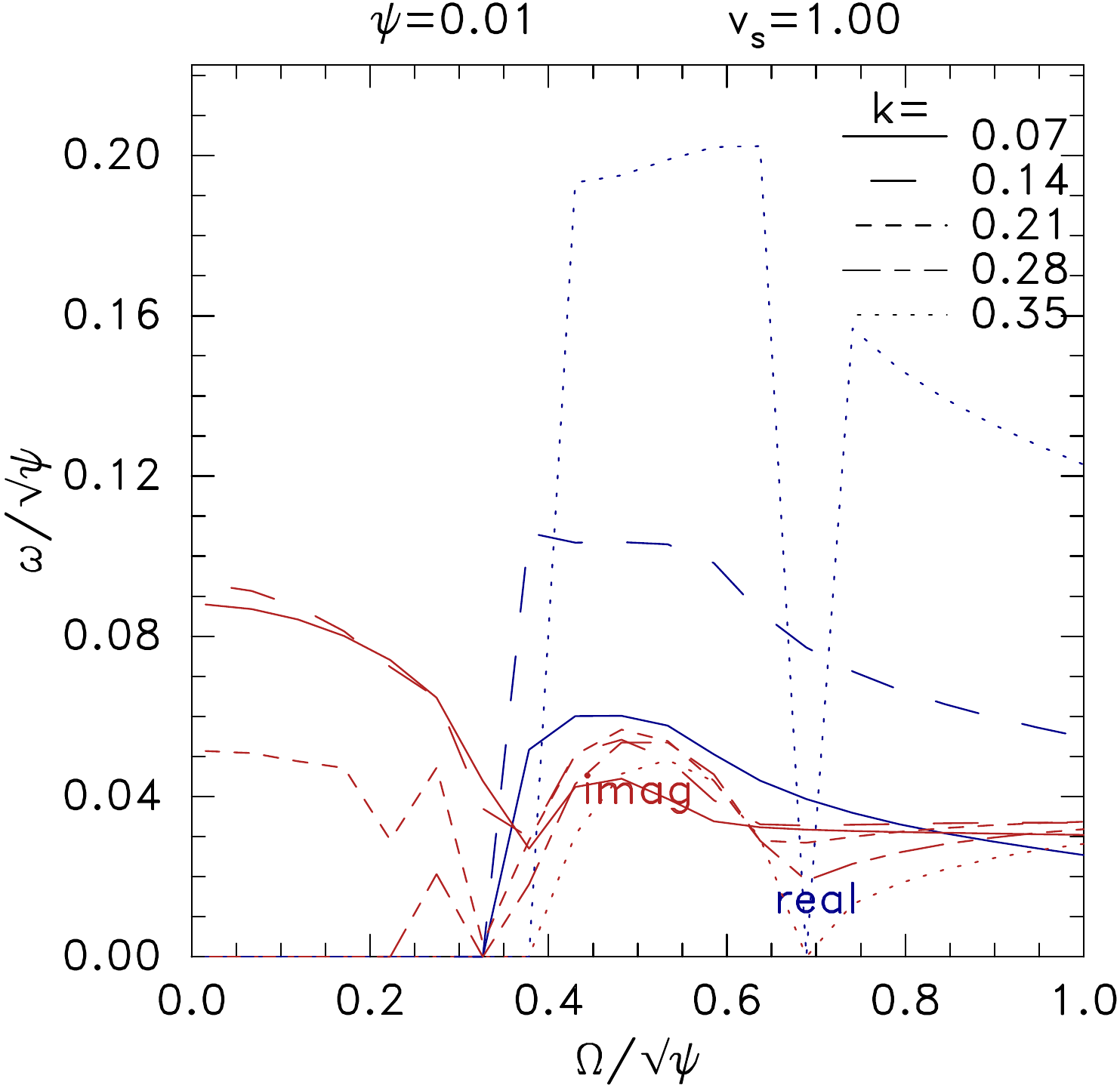}
  \ (b)\hskip-1.5em\includegraphics[width=0.49\hsize]{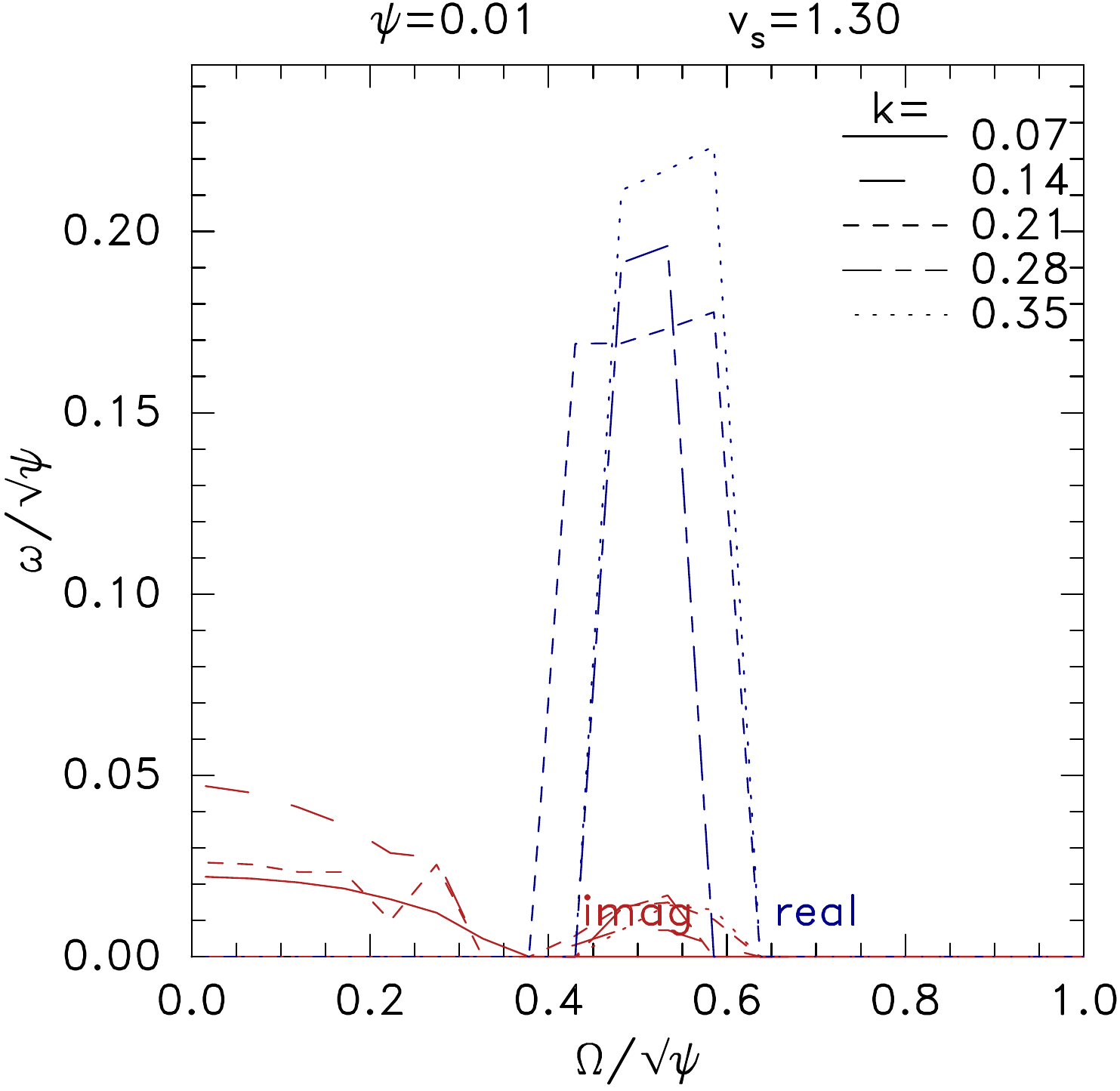}
  \caption{\label{ovo2}As in Fig.\ \ref{ovo1}, except for smaller
    potential amplitude $\psi=0.01$.}
\end{figure}
At smaller potential amplitude, Fig.\ \ref{ovo2}, the double humped
$v_s=1.3$ case (b) looks very different. It is stable for
$\Omega/\sqrt{\psi}\gtrsim0.65$ for all $k$ values, consistent with Fig.\
\ref{omegaofv1}. An oscillatory instability of low growth rate exists
in the range $0.4\lesssim\Omega/\sqrt{\psi}\lesssim0.65$ except at very low
$k$, and a purely growing mode below 0.4 for several $k$ values. In
fact this plot looks extremely similar to Figure 3 of
 \citep{Hutchinson2019}, concerning transverse instability of
\emph{fast} electron holes with zero ion force. That is a sign that
this instability is driven by the electrons with little contribution
from the ions. It is effectively a slightly modified fast hole
transverse instability. Fig.\ \ref{ovo2}(a) with single humped ion
distribution, by contrast, has unstable oscillatory solutions all the
way out to large $\Omega$. These are driven by the combined electron
and ion response, and are limited to slow holes. The low $\Omega$ part
of this plot has a character more like Fig.\ \ref{ovo2}(a) than
\ref{ovo1}(b), and might be thought of as the transverse instability
modified by the slow hole ion response.

\section{Summary}

Slow electron holes are subject to instability of their velocity
because of interaction with bulk ions. Purely growing one-dimensional
($k=0$) instability disappears when a local minimum exists in the ion
velocity distribution ($v_s\gtrsim 1.1v_{ti}$ for symmetric Maxwellian
ion beams), in accordance with earlier studies.  However, an
oscillatory instability (overstability) remains up to relatively deep
local minima, depending on the potential amplitude $\psi$ of the
equilibrium, as shown in Fig.\ \ref{omegaofv1}. Its growth rate is
much smaller than the purely growing mode.
If the peak hole potential $\psi$ is less than
approximately $0.01T_e/e$ (for $T_i=T_e$ but larger for larger $T_i$
see Fig.\ \ref{omegaofTi}), the overstability disappears, this favors
stability of \emph{low amplitude} slow electron holes, which seems
consistent with several observational studies \citep{Kamaletdinov2021}.

When, $k\not=0$, instabilities of slow holes have generally lower
growth rate than when $k=0$, except that when
$(\Omega/\omega_{pe})/\sqrt{e\psi/T_e}\lesssim 0.6$ even low amplitude
holes experience transverse instability driven predominantly by the
electron response and not much affected by ion coupling. Its real
frequency becomes zero if
$(\Omega/\omega_{pe})/\sqrt{e\psi/T_e}\lesssim 0.4$.

An overall caveat to the present work is that it has analyzed only the
pure shift mode. Under some circumstances a distorted shift can become
more unstable \citep{Hutchinson2019}. This might modify the exact values
of the instability thresholds.

The observational implications of the current results are broadly
these.  Slow electron holes in single humped background ion
distributions will speed up without oscillations with growth time of
order $0.3/(\psi^{1/2}\omega_{pe})$. They are predicted to remain slow
for only a few growth times. By contrast, when $f_i$ has a local
minimum in which the hole speed lies, the oscillatory instability
growth time is longer by of order a factor ten; so they remain slow
at least that much longer, The oscillation real angular frequency is of order
$0.1\psi^{1/2}\omega_{pe}$. When the hole potential is smaller or the
$f_i$ local minimum deeper, the growth rate of one-dimensional modes
eventually becomes zero, unless the magnetic field is low enough that
the electron transverse instability is excited. Therefore the
oscillatory instability discovered here can be avoided entirely in
plausible parameter regimes. Furthermore, slow holes might be observed
soon after formation even in regimes that have slowly growing
instabilities.

It would be of considerable interest to verify by simulation the
results of the present study. The required simulations for detailed
quantitative comparison are rather difficult because they need a fully
self-consistent steady slow hole initial equilibrium. My preliminary
efforts, using hole potentials $\psi$ considerably larger (0.2-0.4)
than those for which this analysis predicts stability, do show
oscillatory instabilities which are broadly consistent, being shift
modes in approximately the right frequency range:
$\Real(\omega)/\psi^{1/2}\sim0.1/\omega_{pe}$. However, the
computational cost of simulations with sufficient signal-to-noise or
velocity resolution increases rapidly with decreasing $\psi$, and have
not yet been carried out in the lower-$\psi$ stabilized regime.

\section*{Acknowledgements}
I thank Ivan Vasko and his group for sharing their analysis of
satellite data, which was a major impetus for this study. The work was
not supported by any explicit funding. The codes that calculated the
figures are publically available\footnote{https://github.com/ihutch/shiftmode}.

\bibliography{JabRef}

\end{document}